\documentclass[superscriptaddress, aps, prl, twocolumn]{revtex4-2}

\usepackage{xr-hyper}
\usepackage{dsfont}
\usepackage{upgreek}
\usepackage{amsfonts}
\usepackage{amssymb}
\usepackage{amsmath}
\usepackage{graphicx}
\usepackage{dcolumn}
\usepackage{bm}
\usepackage[colorlinks]{hyperref}
\hypersetup{allcolors= {blue}}
\usepackage{braket}
\usepackage{soul}
\usepackage[most]{tcolorbox}
\usetikzlibrary{patterns}
\pgfdeclarepatternformonly{mystrikeout}{\pgfqpoint{-1pt}{-1pt}}{\pgfqpoint{11pt}{11pt}}{\pgfqpoint{10pt}{10pt}}%
{
  \pgfsetlinewidth{0.4pt}
  \pgfpathmoveto{\pgfqpoint{0pt}{0pt}}
  \pgfpathlineto{\pgfqpoint{10.1pt}{10.1pt}}
  \pgfusepath{stroke}
}
\newtcolorbox{tcbstrikeout}{breakable,
 enhanced jigsaw,
 opacityback=0,
 parbox=false,
 boxrule=0mm,
 top=0mm,bottom=0pt,left=0pt,right=0pt,
 boxsep=0pt,
 frame hidden,
 finish={\fill[pattern=mystrikeout] (frame.north west) rectangle (frame.south east);}
}

\usepackage{tcolorbox}

\usepackage{enumitem}

\begin{document}

\title{Weak Value Advantage in Overcoming Noise}

\newcommand{\bina}{Faculty of Engineering and the Institute of Nanotechnology and Advanced Materials, Bar Ilan University, Ramat Gan, Israel}
\newcommand{\huji}{School of Computer Science and Engineering, Hebrew University, Jerusalem, Israel}

\author{Zohar Schwartzman-Nowik}
\email{zohar.nowik@mail.huji.ac.il}
\affiliation{\huji}
\affiliation{\bina}

\author{Dorit Aharonov}
\affiliation{\huji}

\author{Eliahu Cohen}
\email{eliahu.cohen@biu.ac.il}
\affiliation{\bina}

\date{\today}

\begin{abstract}

The weak value exhibits numerous intriguing characteristics, such as values outside the operator spectrum, leading to unexpected phenomena. 
Nevertheless, the measurement protocol used for measuring the weak value has been the subject of an on-going controversy. In particular, the possibility of gaining a metrological advantage using weak measurements was questioned. A rigorous characterization of this advantage when the primary system is  noisy is still missing. We thus consider here the challenge of learning an unknown operator under the influence of noise on the primary system which could lead to bias in the results. For amplitude and phase damping noise channels, we prove that the weak value measurement protocol (WVMP) 
eliminates the bias to linear order, and this cannot be done with strong measurements. 
Since the WVMP makes use both of weak entanglement as well as postselection, one might suspect that the advantage is solely due to the postselection aspect of the WVMP.
We prove that this is not the case, and that the same advantage of the WVMP is kept even over strong measurement protocols that are allowed to apply postselection.
By this we rigorously prove for the first time the existence of settings in which the WVMP possesses a strict advantage in robustness to noise, even over strong measurements augemented with postselection. 
However, for some noise channels, we show that no advantage is exhibited once both measurement regimes are equipped with postselection.

\end{abstract}

\maketitle

\section{Introduction}
For any operator $A$, initial state $|\psi_{s}\rangle$ and final
state $|\psi_{f}\rangle$, Y. Aharonov, D. Albert, and L. Vaidman defined the weak value (WV) as $A_{w}=\left\langle \psi_{f}|A|\psi_{s}\right\rangle/\left\langle \psi_{f}|\psi_{s}\right\rangle$ \cite{aharonov_how_1988}.
They also constructed a protocol for measuring the WV, which utilizes
both weak measurements as well as postselection. The protocol
includes the primary system, in many cases a photon, as well as an
ancillary probe system, the needle of a measurement device which often lives in the infinite dimensional position space. We will refer to this as the weak value measurement protocol (WVMP).

Since their introduction, the WV and the WVMP have contributed to both
fundamental and applied quantum physics. From the practical point of view, one of the main applications of the WV is the enhancement of high precision measurements \cite{hosten_observation_2008,dixon_ultrasensitive_2009,pfeifer_weak_2011,turner_picoradian_2011, hallaji2017weak, jordan_technical_2014, harris2017weak, chen_beating_2021, martinez2021theory}.
These results triggered a heated debate whether the WV is
advantageous in different scenarios, and specifically when noise is present. 
The vast majority of prior work has studied the case of noise acting
on the probe rather than on the primary measured system,
with few exceptions, such as \cite{ban2013weak, abe2015decoherence, ferraz2024relevance} that focus on the ability of WVs to obtain anomalous values (and not on the advantage of the WVMP compared to other measurement scenarios). 
Some of these works propose evidence
that supports the advantage of the WV and WVMP
\cite{chen_beating_2021,jordan_technical_2014,harris2017weak, feizpour_amplifying_2011,kedem_using_2012,brunner_measuring_2010, pang_protecting_2016, nishizawa_weak-value_2012, AlvesWeakValue},
and some oppose it \cite{ferrie_weak_2014, knee_when_2014}, leaving the general question of the WV advantage in the context of noise open for debate \cite{vaidman_comment_2014,vaidman_weak_2017-1,sinclair2017weak,lyons2015power,wang2016experimental}. 
In this Letter, we attempt to achieve substantial progress towards a clarification of this question. 

\begin{figure}
    \centering
    \includegraphics[width=0.85\linewidth]{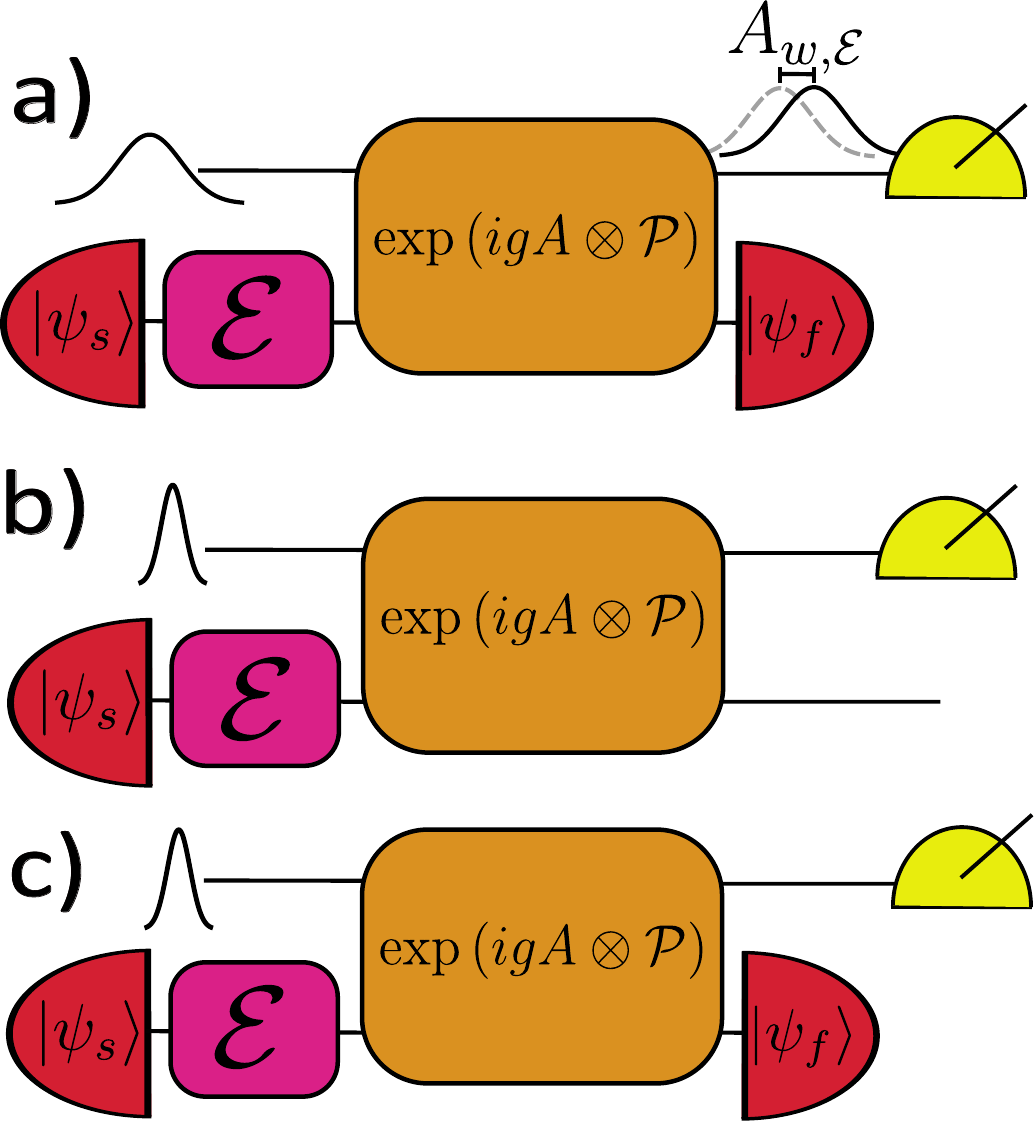}
    \caption{a. The WVMP of the WV,
consisting of a pre-selected state $|\psi_s\rangle$, noise $\mathcal{E}$, weak entanglement $\exp\left({igA\otimes \mathcal{P}}\right)$ and postselection $|\psi_f\rangle$. b. The strong measurement protocol without postselection, c. The strong measurement protocol with postselection. The entanglement is weak (strong) if the standard deviation of the probe state is large (small) compared to the interaction strength $g$.}
    \label{fig:fig1}
\end{figure}

Before proceeding to our results, we recall the details of the WVMP.
The WVMP consists of the following steps:
\begin{enumerate}
\item Initialize the primary system in state $|\psi_{s}\rangle$ and initialize
the probe system to a Gaussian of variance $\Delta^2$ centered around position $q=0$, given
by $\int dq\frac{1}{\left(2\pi\Delta^{2}\right)^{\frac{1}{4}}}\exp\left(-\frac{q^{2}}{4\Delta^{2}}\right)|q\rangle$.
\item Weakly couple the two systems by applying the interaction Hamiltonian
given by $H=\tilde{g}\left(t\right)A\otimes\mathcal{P}$, where $\mathcal{P}$
is the momentum operator of the probe, for time $T$ for which
$\int_{0}^{T}\tilde{g}\left(t\right)dt\equiv g\ll\Delta$.
\item Measure the primary system and postselect on a final state 
$|\psi_{f}\rangle$ which is not orthogonal to  $|\psi_s\rangle$.
\item Measure the probe system in the position basis.
\end{enumerate}
In  \cite{aharonov_how_1988} the authors showed that if $\frac{g}{\Delta}\ll\left|A_{w}\right|^{-1}$
and $\frac{g}{\Delta}\ll\left|\frac{\left\langle \psi_{f}|A^{n}|\psi_{s}\right\rangle }{\left\langle \psi_{f}|A|\psi_{s}\right\rangle }\right|^{-\frac{1}{n-1}}$
then the expectation value of the probe measurement is
$g\mathcal{R}\left(A_{w}\right)$, where $\mathcal{R}\left(A_{w}\right)$ is the real part of $A_{w}$, and the
variance of this measurement is proportional to $\Delta^2$ \cite{tamir_introduction_2013}. 
In Fig. \ref{fig:fig1}a we present the WVMP together with a noise channel affecting the primary system.
The WVMP can easily be generalized, as was done in \cite{johansen2004nonclassicality, dressel2012significance},
for initial mixed state $\rho_{s}$, where the WV
becomes $A_{w}=\frac{\left\langle \psi_{f}|A\rho_{s}|\psi_{f}\right\rangle }{\left\langle \psi_{f}|\rho_{s}|\psi_{f}\right\rangle }$.
The derivation 
can be found in Appendix \ref{app:wv and wvmp derivation}.

\section{Setup and definitions}

In this work we provide a proof of the advantage of the WVMP which is rigorous and qualitative, in the sense that we explicitly define a setting and a task and show that the WVMP can succeed in accomplishing the task, while strong measurements, even with postselection, cannot. 
Informally, We study the task of learning all the parameters of an unknown operator A acting on the primary system in the presence of amplitude and phase damping of magnitude $\gamma$, acting on the primary system. 
We show that in the presence of such noise processes, WVMP can provide a  correct outcome up to first order in $\gamma$, and thus is robust to such noise processes. 
In contrast, we show that any strong measurement, even when augmented with postselection, would incur a bias that grows linearly with $\gamma$ and so strong measurements even with post selection 
do not exhibit robustness to these noise processes.  
The advantage even over strong measurements with postselection goes to show that the source of the advantage is not only due to the postselection \cite{arvidsson-shukur_quantum_2020, lupu-gladstein_negative_2022}, but rather due to the combination of weak entanglement and postselection exhibited in the WVMP \cite{nakamura2012evaluation, tamir_introduction_2013, porotti2022deep, lantano2025unlocking}.
To the best of our knowledge this is the first clear proof of a qualitative advantage of the WVMP over strong measurement protocols, especially over strong measurements augmented with postselection.

In order to state our results we now turn to define the setup, competing measurement protocols, and the noise channels that we are considering.

\emph{Setup.---}
As depicted in Fig. \ref{fig:fig1}a, the WVMP consists of an initial state, a final state, and an entangling operator.
This measurement protocol is implemented for various reasons in various different setups for various different goals. 
In this work we will consider a scenario inspired by the work of Hosten et. al, where the Spin Hall Effect of Light was confirmed using the WVMP \cite{hosten_observation_2008}.
This scenario includes an initial and final state which are fully known and controllable, while it is the entangling operator, which is arbitrary and unknown, that we are trying to learn. 
We consider a primary system of a single qubit, which is not confined to be in a pure state, but can rather be a mixed state as well.
Note that there are other common scenarios as well, such as
where the final state and entangling operator are known, while the initial state is only known up to some unknown parameter, and the goal of the measurement protocol is to assess this parameter.

We consider a scenario where noise introduces a systematic error, leading to a difference between the expectation values with and without noise, termed ``the bias''. 
Notably, while the variance of measurement outcomes 
can be reduced to arbitrarily small values by repeatedly taking measurements, the same does not hold for the bias. 
When noise induces a bias, the error cannot be eliminated even in the limit of an infinite number of samples, making it crucial to overcome these errors. For  this reason, the sampling complexity becomes irrelevant.
In other words, this work is more focused on the accuracy of the measurement outcomes rather than their precision (or variance). As a consequence, common metrics like the Fisher information \cite{jordan_technical_2014} become less suitable and more useful tools are introduced. Eventually, the presented advantage is not only quantitative, but also qualitative -- in some scenarios, the WVMP method succeeds, while projective measurement with (or without) postselection fails.
Also, as can be seen in the results presented later, the states that are used to obtain the advantage are not nearorthogonal states, and so not many samples are needed to obtain successful postselections.

\emph{Strong measurements.---} 
We now present the measurement protocols that we compare to the WVMP. 
The strong measurement is simply a PVM - projection-valued-measure, which is equivalent to a von Neumann measurement, where the possible results of the measurement are the eigenvalues of the operator that is measured, and as a result, the initial state collapses to the corresponding eigenstate, with a probability corresponding to the overlap between the initial state and eigenstate. In order to facilitate the comparison to the WVMP, we show in Appendix \ref{app: derivation strong measurement} that this is also equivalent to a modified WVMP setup, but with small probe variance compared to a large coupling constant, i.e. $\frac{g}{\Delta}$ is large, and no postselection, as depicted in Fig. \ref{fig:fig1}b. 
As $\frac{g}{\Delta}$ increases in magnitude, this approach increasingly approximates the von Neumann projective measurement, as the expectation value of such a measurement is $g\left\langle \psi_{s}|A|\psi_{s}\right\rangle $
and the variance is $g^{2}\left(\left(\left\langle \psi_{s}|A^{2}|\psi_{s}\right\rangle -\left\langle \psi_{s}|A|\psi_{s}\right\rangle ^{2}\right)+\left(\frac{\Delta}{g}\right)^{2}\right).$ The full derivation can be found in Appendix \ref{app: derivation strong measurement}.

We also compare to a stronge measurement with postselection, which is a strong measurement as described above, followed by a postselection.
In section II of the appendix we show that this is equivalent to the WVMP with large $\frac{g}{\Delta}$ and with postselection, as depicted in Fig. \ref{fig:fig1}c.

\emph{The noise channels.---}
We consider noise acting on the primary system, as can be seen in Fig. \ref{fig:fig1}. There are many different noise channels that can act on the primary
system. A general channel can be represented by Kraus operators as
$\mathcal{E}\left(\rho\right)=\sum_{k}E_{k}\rho E_{k}^{\dagger}$
with $\sum_{k}E_{k}^{\dagger}E_{k}=I$. 
In this paper we will use the term "noise channel" for a {\it family}
of noise channels, parameterized by a noise parameter $\gamma\in\left[0,1\right]$
such that when $\gamma=0$ the channel is the identity channel. 
For the sake of brevity, we will frequently omit the term ``parameterized'' as well as the parameter $\gamma$ when referring to such noise channels. 
One example
is the Pauli noise channel, 
\begin{equation}
\mathcal{E}_{\mathcal{P}}\left(\rho\right)=\left(1-\gamma\right)\rho+\gamma\sum_{\sigma\in P^{n},\sigma\neq I}\lambda_{\sigma}\sigma\rho\sigma,\label{eq:pauli noise channel}
\end{equation}
where $\lambda_{\sigma}$ are unknown, 
$\sum_{\sigma}\lambda_{\sigma}=1$ and $P^{n}$ is the Pauli group
on $n$ qubits. 
Other examples are the amplitude or phase damping channel on a single qubit where $\mathcal{E}_{D}\left(\rho\right)=E_{0}\rho E_{0}^{\dagger}+E_{1}\rho E_{1}^{\dagger}$ for  $E_{0}=\left(\begin{matrix}1 & 0\\
0 & \sqrt{1-\gamma}
\end{matrix}\right)$ for both amplitude and phase damping and $E_{1,AD}=\left(\begin{matrix}0 & \sqrt{\gamma}\\
0 & 0
\end{matrix}\right)$, while 
$E_{1,PD}=\left(\begin{matrix}0 & 0\\
0 & \sqrt{\gamma}
\end{matrix}\right)$ for phase damping. 

A standard way to describe noise acting during the process of the WVMP is through a Master equation in the Lindblad form. 
This equation encompasses both the intentional WVMP and the accompanying noise terms.
In Appendix \ref{app: lindblad approximation} we show that when both the noise parameter $\gamma$ and the entanglement parameter $g$ 
are small, we can approximate this process by a simpler noise model, where noise acts on the initial state followed by ideal and noiseless entangling interactions.
In this Letter, we will work with this simplified noise model as it offers a more straightforward analytical approach and closely approximates the more complex noise model in many cases. 

For such noise the initial state changes to $\mathcal{E}\left(\rho_s\right)$ and so the expected value of the probe shifts by 

\begin{equation}
A_{w,\mathcal{E}}=\frac{\left\langle \psi_{f}|A\mathcal{E}\left(\rho_s\right)|\psi_{f}\right\rangle }{\left\langle \psi_{f}|\mathcal{E}\left(\rho_s\right)|\psi_{f}\right\rangle }.
\label{eq:noisy weakval}
\end{equation}

We can expand $A_{w,\mathcal{E}}$ as a Taylor series in $\gamma$ to obtain
\begin{equation}
A_{w,\mathcal{E}}=A_{w}+\gamma\Delta_{\mathcal{E}}+O\left(\gamma^{2}\right),
\label{eq:taylor of noisy weakval}
\end{equation}
where we define
$\gamma\Delta_{\mathcal{E}}$ as the bias of the WV for the noise
channel $\mathcal{E}$ to first order in $\gamma$. In this Letter, we aim to identify scenarios in which $A_{w,\mathcal{E}}$ equals $A_w$ to first order in $\gamma$, which is the case when $\Delta_{\mathcal{E}}=0$.
We will assert that there is a noise sensitivity advantage for the WVMP compared to strong measurement (resp., with postselection) in situations where $\Delta_{\mathcal{E}}=0$
while it is impossible to eliminate the first order in $\gamma$ in the expectation value of strong measurements (resp., with postselection). More formally, we define:

\emph{Definition - Noise sensitivity advantage for the WVMP}:
For a noise channel $\mathcal{E}$,
\begin{enumerate}
    \item {
    There is an advantage compared to strong measurements if the following conditions hold:
    \begin{enumerate}
        \item {By using the WVMP it is possible to estimate all elements of $A$, i.e. $a_{ij}$ for all $i,j$, under the noise channel $\mathcal{E}$, with bias $O\left(\gamma^{2}\right)$}
        \item{For any protocol which uses expectation values of strong measurements only, $\exists i,j$ for which, under the noise channel $\mathcal{E}$, $a_{ij}$ can only be estimated with bias of linear order in $\gamma$.}
    \end{enumerate}
    }
    \item {There is an advantage compared to strong measurements with postselection if also:}
    \begin{enumerate}[start=3]
        \item {For any protocol which uses strong measurements and postselection, $\exists i,j$ for which, under the noise channel $\mathcal{E}$, $a_{ij}$ can only be estimated with bias of linear order in $\gamma$.}
    \end{enumerate}
\end{enumerate}

\section{Results}

\emph{Theorem.---} (Advantage for amplitude and phase damping noise compared to strong measurements, even with postselection) When
$A$ is an unknown Hermitian operator acting on a primary system of
a single qubit which suffers from a combined channel of amplitude and phase damping noise, or amplitude damping alone, then the WVMP has a noise sensitivity advantage compared to strong measurements as well as strong measurements with postselection.

\emph{Proof---} 
There are three parts to proving the theorem. Claim 1 - the WVMP can succeed in the task. Claim 2 - strong measurements cannot succeed in the task. Claim 3 - strong measurements with postselection cannot succeed in the task.
We will therefore prove all three of these claims in order to prove the theorem.
We present here the main ideas of the proof, while all the details are presented in Appendix \ref{app:main theormen proof}.

First we observe that for a combined noise channel $\mathcal{E}_{2}\circ\mathcal{E}_{1}$
where the noise parameter of $\mathcal{E}_{i}$ is $\lambda_{i}\gamma$,
and $\gamma$ is the noise parameter for the combined noise channel,
then $A_{w,\mathcal{E}_{2}\circ\mathcal{E}_{1}}=A_{w}+\gamma\lambda_{1}\Delta_{\mathcal{E}_{1}}+\gamma\lambda_{2}\Delta_{\mathcal{E}_{2}}+O(\gamma^2)$.
Hence, whenever the linear order vanishes for both of the separate
channels, i.e. $\Delta_{\mathcal{E}_{1}}=\Delta_{\mathcal{E}_{2}}=0$, it will also vanish for the combined channel. 

\emph{Proof of Claim 1, WVMP can accomplish the task---}
By substituting an amplitude and phase damping as the noise channel in Eq. (\ref{eq:noisy weakval}) and expanding the Taylor series as in Eq. (\ref{eq:taylor of noisy weakval}) we find that the linear order of the noise for amplitude and phase damping noise is given by 
\begin{equation}
    \Delta_{\text{\text{amplitude damping}}}=\frac{\left\langle f|AM^{\text{ad}}|f\right\rangle }{\left\langle f|\rho|f\right\rangle }-\frac{\left\langle f|M^{\text{ad}}|f\right\rangle }{\left\langle f|\rho|f\right\rangle }\frac{\left\langle f|A\rho|f\right\rangle }{\left\langle f|\rho|f\right\rangle },
\end{equation}

where we define $M^{\text{ad}}=\left(\begin{matrix}\rho_{22} & -\frac{1}{2}\rho_{12}\\
-\frac{1}{2}\rho_{21} & -\rho_{22}
\end{matrix}\right)$.

There are three families of states for which this linear order vanishes for any operator $A$.

\begin{equation}
|\psi_{f}\rangle=\left(\begin{matrix}1\\
0
\end{matrix}\right),\rho=\left(\begin{matrix}\rho_{11} & 0\\
0 & 1-\rho_{11}
\end{matrix}\right)\longrightarrow A_{w}=a_{11},
\end{equation}

\begin{equation}
|\psi_{f}\rangle=\left(\begin{matrix}0\\
1
\end{matrix}\right),\rho=\left(\begin{matrix}\rho_{11} & 0\\
0 & 1-\rho_{11}
\end{matrix}\right)\longrightarrow A_{w}=a_{22},
\end{equation}

\begin{equation}
|\psi_{f}\rangle=\left(\begin{matrix}f_{1}\\
f_{2}
\end{matrix}\right),\rho=\left(\begin{matrix}1 & 0\\
0 & 0
\end{matrix}\right)\longrightarrow A_{w}=a_{11}+a_{21}\frac{f_{2}^{*}}{f_{1}^{*}}.
\end{equation}

These states can be used to fully learn $A$ as follows. $a_{11}$ and $a_{22}$ are the direct results of the WV for the first two families of solutions. Since the shift in the probe is the real part of the WV, choosing $f_1=f_2=\frac{1}{\sqrt{2}}$, we can learn the real part of $a_{12}$, and choosing $f_{1}=\frac{1}{\sqrt{2}}$ and $f_{2}=\frac{i}{\sqrt{2}}$ allows learning the imaginary part of $a_{12}$. All these are free of linear order of the noise since the linear order vanishes on these sets of states. 

And so in total we showed that we can use the WVMP in order to fully learn $A$ without effect of the linear order of the noise, thus concluding the proof of Claim 1.

\emph{Proof of Claim 2, strong measurements cannot accomplish the task---}
In fact, Claim 2 follows from Claim 3. However, for didactic reasons, we have chosen to present the proof of Claim 2 separately.
For strong measurements 
$\left\langle Q\right\rangle =\gamma \text{Tr} \left(A\mathcal{E}\left(\rho\right)\right)$. Substituting amplitude damping and expanding the Taylor series we find that the only state for which the linear order vanishes for any operator $A$ is $\rho=\left(\begin{matrix}1 & 0\\0 & 0 \end{matrix}\right)$ in which case $\text{Tr}\left(A\rho\right)=a_{11}$. And so no other information about $A$ can be obtained without linear order affect of the noise, proving that the strong measurement cannot succeed in the task, concluding the proof of the second claim.

\emph{Proof of Claim 3, strong measurements with postselection cannot accomplish the task---}
In order to prove Claim 3 we will start by proving a lemma. 

Lemma - the result of the strong measurement with postselection is not affected by the noise in linear
order only if the initial state is not affected by the noise in linear order.

This is opposed to the case of the WVMP, where certain combinations of initial and final states give rise to WVs that overcome the noise, even if the initial state alone does suffer significantly from the noise. 
The proof of the lemma is rather technical and so is presented in the appendix.

In order to complete the proof of Claim 3, we must find the initial states that are not affected by amplitude damping to linear order. 
The state 
$\rho=\left(\begin{matrix}1 & 0\\0 & 0 \end{matrix}\right)$ is the only state satisfying this condition.
Lastly, even if we restrict ourselves to $A$ of the sort 
$A=a_{0}|a_{0}\rangle\langle a_{0}|+a_{1}|a_{1}\rangle\langle a_{1}|=\left(\begin{matrix}a_{0} & 0\\
0 & a_{1}
\end{matrix}\right)$
we get
$\left\langle Q\right\rangle _{strong\,with\,post}=a_{0}\left|f_{0}\right|^{2}$,
and so we cannot learn $a_1$ and so we cannot learn $A$ even in this restricted case, completing the proof of Claim 3.
$\blacksquare$

The above results prove, for the first time, that the WVMP provides a strict advantage in terms of robustness to noise, even over strong measurements with postselection, by defining a task that the WVMP can accomplish, while strong measurements, even with postselection, cannot.

\emph{Additional results.---} 
In Appendices \ref{app: additional theorem 1} and \ref{app: additional theorem 2} we present additional results. We found that for a Pauli noise channel the WVMP has an advantage over strong measurements but not over strong measurements with postselection. 
Similarly, for a unital noise channel, there is only advantage compared to strong measurements without postselection.
These results show that the advantage of the WVMP is not absolute, and there are some cases with more advantage than others, creating a hierarchy of noise channels which manifest different levels of advantage for the WVMP.

Not only different noise channels affect the level of advantage. The current work sheds some light on the sets of initial and final states that can provide such advantages.
Interestingly, the maximally mixed state turns out to play an important role here. In some cases, it is the {\it only} state that can yield an advantage for the WVMP. We prove this is the case for the noise being a probabilistic unitary channel \emph{$\mathcal{E}_{\text{U}}\left(\rho\right)=\left(1-p\right)\rho+pU\rho U^{\dagger}$,}
for some fixed unknown unitary matrix $U$. 
We show that a maximally mixed initial state leads to an advantage for the WVMP over strong measurements for this channel.
We show that for any other initial state 
there will always be a unitary $U$ for which the requirement  $\Delta_{\mathcal{E}_{\text{U}}}=0$ does not hold. 
Using Weingarten functions, we further extend this result to the case in which one wants to achieve an advantage not for all fixed unitaries but for most such unitaries. 
The proofs are presented in Appendix \ref{app:maximally mixed}. 
Another evidence for the fact that incoherent probes and even the maximally mixed state can be useful in metrology is presented in \cite{goldberg_evading_2023}, where the authors use superpositions of causal order for multiparameter estimation.

\section{Conclusion and open questions}

We have demonstrated the utility of the WVMP in effectively overcoming noise channels that affect the primary system. 
Our findings showcase that learning the operator $A$ which governs the entanglement between the probe and the primary system under the influence of an amplitude and phase damping channel, can be accomplished successfully with an impact that is quadratically better when using WVMP compared to any method using strong measurements alone, and also compared to any method using strong measurements and postselection. In doing so, we have underscored the benefit of the WVMP and WV, particularly in the less explored scenario of noise affecting the primary system.
We note that the advantage of the WVMP lies in its ability to cancel the linear order contribution of noise in the measurement result, even when the initial state is linearly affected by noise, an ability that the strong measurement lacks

In contrast, we prove that for different types of noise channels, specifically Pauli or unital noise, the advantage of the WVMP disappears once the strong measurement is supplemented with postselection, indicating that this advantage has its limits.

Hopefully, this work will lead to extensions to more general cases, and in particular to the interesting scenario where the noise acts both on the system and on the probe, as well as to multi-particle systems or higher dimensional systems.
Another question worth exploring is what happens under general noise channels:
Is there an advantage for the WVMP when we have no guarantee regarding the nature of the noise channel?
Lastly, it would be instructive to take into account the variance and sampling overhead associated with the WVMP, 
and find a trade-off between the variance and the bias. This exploration could provide insights into the practical limitations and trade-offs in applying the WVMP in different settings and applications.

\section*{Acknowledgments}
We  thank Ronnie Kosloff and Raam Uzdin for insightful discussions.

\bibliography{citations}

\appendix
\onecolumngrid
\setcounter{secnumdepth}{3} 

\section*{Appendix}
The Appendix is organized as follows.
In Appendix \ref{app:wv and wvmp derivation}, we derive the weak value for different cases involving pure and mixed initial and final states.
In Appendix \ref{app: derivation strong measurement}, we derive the corresponding values in the strong measurement limit and compare them to the weak values discussed in the main text.
In Appendix \ref{app: lindblad approximation} we derive of the approximation of a Lindblad Master equation of the combined weak measurement protocol (WMP) and noise by a Kraus noise channel followed by an ideal WMP.
In Appendix \ref{app:main theormen proof}, we provide the proof of the main theorem in the paper, while Appendices \ref{app: additional theorem 1} and \ref{app: additional theorem 2}, present two additional theorems and their proofs, corresponding to cases where the advantage of the WMP is less pronounced.
Finally, in Appendix \ref{app:maximally mixed}, we analyze the uniqueness of the maximally mixed state.

\section{The weak measurement protocol of the weak value \label{app:wv and wvmp derivation}}

For any operator $A$, initial state $|\psi_{s}\rangle$ and final (non-orthogonal)
state $|\psi_{f}\rangle$, the weak value is defined as
\begin{equation}
A_{w}=\frac{\left\langle \psi_{f}|A|\psi_{s}\right\rangle }{\left\langle \psi_{f}|\psi_{s}\right\rangle }.
\end{equation}

We will show that if the initial state $\rho_{s}$ is not pure, this generalizes to 
\begin{equation}
A_{w}=\frac{\left\langle \psi_{f}|A\rho_{s}|\psi_{f}\right\rangle }{\left\langle \psi_{f}|\rho_{s}|\psi_{f}\right\rangle }.
\end{equation}

If also the final state $\rho_{f}$ is not pure, this generalizes
to 
\begin{equation}
A_{w}=\frac{\text{Tr}\left(\rho_{f}A\rho_{s}\right)}{\text{Tr}\left(\rho_{f}\rho_{s}\right)}.
\end{equation}

Aharonov, Albert, and Vaidman constructed a protocol for measuring
the weak value, which we will denote as the weak measurement protocol (WMP),
and utilizes both weak measurements as well as postselection. The
WMP goes as follows (the adaptations for mixed
states are straightforward):

\begin{enumerate}
\item Initialize the primary system in state $|\psi_{s}\rangle$ and initialize
the probe system to a Gaussian centered around position $q=0$, given
by $\int dq\frac{1}{\left(2\pi\Delta^{2}\right)^{\frac{1}{4}}}\exp\left(-\frac{q^{2}}{4\Delta^{2}}\right)|q\rangle$.
\item Weakly couple the two systems by applying the interaction Hamiltonian
given by $H=\tilde{g}\left(t\right)A\otimes\mathcal{P}$ where $\mathcal{P}$
is the momentum operator and the coupling parameter $\tilde{g}(t)$ obeys $\int_{0}^{T}g\left(t\right)dt\equiv g\ll\Delta$, throughout the interaction duration $T$.
\item Measure the primary system and postselect on the final state being
$|\psi_{f}\rangle$ which is not orthogonal to the initial state.
\item Measure the probe system in the position basis.
\end{enumerate}

The expectation of the probe measurement is $\mathcal{R}\left(A_{w}\right)$, i.e. the
real part of the weak value. In the following we will show this is
indeed the case. 

\subsection{Weak value for pure states \label{subsec:proof weakval}}

We will denote 
\begin{equation}
\phi\left(q\right)=\left\langle q|\phi\right\rangle =\frac{1}{\left(2\pi\Delta^{2}\right)^{\frac{1}{4}}}\exp\left(-\frac{q^{2}}{4\Delta^{2}}\right),\label{eq:probe initial}
\end{equation}
 and so the initial state of the probe is given by $|\phi\rangle=\int dq\phi\left(q\right)|q\rangle$.
After the initialization, the joint system and probe is given by $|\psi_{s}\rangle\otimes\int dq\phi\left(q\right)|q\rangle$.
After the weak coupling, and taking $\hbar=1$ the joint state is 
\begin{equation}
|\Phi\rangle=e^{-ig A\otimes\mathcal{P}}|\psi_{s}\rangle\otimes|\phi\rangle.\label{eq:before postselection}
\end{equation}
Postselecting the primary system in the final state $|\psi_{f}\rangle$,
we are left with the unnormalized probe state

\begin{align}
|\phi_{f}\rangle & =\left\langle \psi_{f}|e^{-ig A\otimes\mathcal{P}}|\psi_{s}\right\rangle \otimes|\phi\rangle\\
 & \simeq\left\langle \psi_{f}|I\otimes I-ig A\otimes\mathcal{P}|\psi_{s}\right\rangle \otimes|\phi\rangle\label{eq:approx1}\\
 & =\left\langle \psi_{f}|\psi_{s}\right\rangle \left(1-ig A_{w}\mathcal{P}\right)|\phi\rangle\\
 & \simeq\left\langle \psi_{f}|\psi_{s}\right\rangle \exp\left(-ig A_{w}\mathcal{P}\right)|\phi\rangle.\label{eq:approx2}
\end{align}
These approximations hold when $\frac{\left|g\right|}{\Delta}\left|\frac{\left\langle \psi_{f}|A^{n}|\psi_{s}\right\rangle }{\left\langle \psi_{f}|A|\psi_{s}\right\rangle }\right|^{\frac{1}{n-1}}\ll1$
and $\frac{\left|g A_{w}\right|}{\Delta}\ll1$. Now, since $\mathcal{P}$
is the generator of translations, the unnormalized probe state is
\begin{equation}
|\phi_{f}\rangle=\left\langle \psi_{f}|\psi_{s}\right\rangle |\phi\left(q-g A_{w}\right)\rangle=\left\langle \psi_{f}|\psi_{s}\right\rangle \frac{1}{\left(2\pi\Delta^{2}\right)^{\frac{1}{4}}}\int dq\exp\left(-\frac{\left(q-g A_{w}\right)^{2}}{4\Delta^{2}}\right)|q\rangle,
\end{equation}

with $\left\langle \phi_{f}|\phi_{f}\right\rangle =\left|\left\langle \psi_{f}|\psi_{s}\right\rangle \right|^{2}$.
Now, the expectation value of probe position $Q$ is:

\begin{equation}
\mathbb{E}\left(Q\right)=\frac{\left\langle \phi_{f}|Q|\phi_{f}\right\rangle }{\left\langle \phi_{f}|\phi_{f}\right\rangle }=\frac{1}{\sqrt{2\pi\Delta^{2}}}\int dq\cdot q\exp\left(-\frac{\left(q-g\mathcal{R}\left(A_{w}\right)\right)^{2}}{2\Delta^{2}}\right)=\frac{1}{\sqrt{2\pi\Delta^{2}}}\sqrt{2\pi\Delta^{2}}g\mathcal{R}\left(A_{w}\right)=g\mathcal{R}\left(A_{w}\right).
\end{equation}

And since $\frac{\left\langle \phi_{f}|Q^{2}|\phi_{f}\right\rangle }{\left\langle \phi_{f}|\phi_{f}\right\rangle }=\frac{1}{\sqrt{2\pi\Delta^{2}}}\int dq\cdot q^{2}\exp\left(-\frac{\left(q-g\mathcal{R}\left(A_{w}\right)\right)^{2}}{2\Delta^{2}}\right)=g^{2}\mathcal{R}\left(A_{w}\right)^{2}+\Delta^{2}$,
the variance is

\[
\text{Var}\left(Q\right)_{\phi_{f}}=\left\langle \phi_{f}|Q^{2}|\phi_{f}\right\rangle -\left\langle \phi_{f}|Q|\phi_{f}\right\rangle ^{2}=g^{2}\mathcal{R}\left(A_{w}\right)^{2}+\Delta^{2}-g^{2}\mathcal{R}\left(A_{w}\right)^{2}=\Delta^{2}.
\]

$\square$

So small $\Delta$ will result in a small variance around the weak
value, but on the other hand, the approximations done in Eqs. (\ref{eq:approx1})
and (\ref{eq:approx2}) hold for large $\frac{\Delta}{g}$, and so there is
a trade-off between the accuracy of the approximation and the variance
of the measurement, and an optimal $\Delta$ can be chosen by the
requirements of the problem at hand.

\subsection{Weak value for mixed initial state \label{subsec:weakval mixed}}

We define the initial state of the probe and the interaction Hamiltonian
in the same way as above, but the initial state of the system is now
the mixed state $\rho_{s}$, and so the joint initial state is $\rho_{s}\otimes|\phi\rangle\langle\phi|$.
After implying the interaction Hamiltonian we have

\begin{equation}
\exp\left(-ig A\otimes P\right)\left(\rho_{s}\otimes|\phi\rangle\langle\phi|\right)\exp\left(ig A\otimes\mathcal{P}\right).\label{eq:mixed before postselection}
\end{equation}

Applying post selection on $\Pi_{f}=|\psi_{f}\rangle\langle\psi_{f}|$:
the un-normalized resulting state is:

\begin{align}
\rho_{f} & =\text{Tr}\left(|\psi_{f}\rangle\langle\psi_{f}|\left(\exp\left(-ig A\otimes\mathcal{P}\right)\left(\rho_{s}\otimes|\phi\rangle\langle\phi|\right)\exp\left(ig A\otimes\mathcal{P}\right)\right)\right)\\
 & =\left\langle \psi_{f}|\exp\left(-ig A\otimes\mathcal{P}\right)\left(\rho_{s}\otimes|\phi\rangle\langle\phi|\right)\exp\left(ig A\otimes\mathcal{P}\right)|\psi_{f}\right\rangle \label{eq:mixed post selection}\\
 & \simeq\left\langle \psi_{f}|\left(I\otimes I-ig A\otimes\mathcal{P}\right)\left(\rho_{s}\otimes|\phi\rangle\langle\phi|\right)\left(I\otimes I+ig A\otimes\mathcal{P}\right)|\psi_{f}\right\rangle \\
 & \simeq\left\langle \psi_{f}|\rho_{s}|\psi_{f}\right\rangle \left(1-ig A_{w}\mathcal{P}\right)|\phi\rangle\langle\phi|\left(1+ig A_{w}\mathcal{P}\right)\\
 & \simeq\left\langle \psi_{f}|\rho_{s}|\psi_{f}\right\rangle e^{-ig A_{w}\mathcal{P}}|\phi\rangle\langle\phi|e^{ig A_{w}^{*}\mathcal{P}}
\end{align}

for 
\begin{equation}
A_{w}=\frac{\left\langle \psi_{f}|A\rho_{s}|\psi_{f}\right\rangle }{\left\langle \psi_{f}|\rho_{s}|\psi_{f}\right\rangle }.\label{eq:mixed initial}
\end{equation}

\subsection{Weak value for mixed initial and final states }

For initial system state $\rho_{s}$, final system state $\rho_{f}$,
the joint initial state is as before

\begin{equation}
\rho_{s}\otimes|\phi\rangle\langle\phi|=\rho_{s}\otimes\left(\int dq\exp\left(-\frac{q^{2}}{4\Delta^{2}}\right)|q\rangle\right)\left(\int dq'\exp\left(-\frac{\left(q'\right)^{2}}{4\Delta^{2}}\right)\langle q'|\right).
\end{equation}

After the weak interaction, as before we have

\begin{equation}
\exp\left(-ig A\otimes\mathcal{P}\right)\left(\rho_{s}\otimes|\phi\rangle\langle\phi|\right)\exp\left(ig A\otimes P\right).
\end{equation}

After postselecting on the final state $\rho_{f}$ the first order in $g$ of the unnormalized
state of the probe is given by

\begin{align}
 & \text{Tr}\left(\rho_{f}\exp\left(-ig A\otimes\mathcal{P}\right)\left(\rho_{s}\otimes|\phi\rangle\langle\phi|\right)\exp\left(ig A\otimes\mathcal{P}\right)\right)\\
\simeq & \text{Tr}\left(\rho_{f}\left(I\otimes I-ig A\otimes\mathcal{P}\right)\left(\rho_{s}\otimes|\phi\rangle\langle\phi|\right)\left(I\otimes I+ig A\otimes\mathcal{P}\right)\right)\\
\simeq & \text{Tr}\left(\rho_{f}\rho_{s}\right)|\phi\rangle\langle\phi|-ig \text{Tr}\left(\rho_{f}A\rho_{s}\right)\mathcal{P}|\phi\rangle\langle\phi|+ig \text{Tr}\left(\rho_{f}\rho_{s}A\right)\otimes|\phi\rangle\langle\phi|\mathcal{P}\\
\simeq & \text{Tr}\left(\rho_{f}\rho_{s}\right)\left(1-ig A_{w}\mathcal{P}\right)|\phi\rangle\langle\phi|\left(1+ig A_{w}^{*}\mathcal{P}\right)
\end{align}

where we define $A_{w}=\frac{\text{Tr}\left(\rho_{f}A\rho_{s}\right)}{\text{Tr}\left(\rho_{f}\rho_{s}\right)}$,
and so $A_{w}^{*}=\frac{\text{Tr}\left(\rho_{f}\rho_{s}A\right)}{\text{Tr}\left(\rho_{f}\rho_{s}\right)}$
since $\rho_{f},A,\rho_{s}$ are all Hermitian.

\section{The limit of strong measurement \label{app: derivation strong measurement}}

We will now show that performing the same protocol, but instead with
no postselection and in the limit of strong measurement, which means
$\Delta\ll g \delta a$ for $\delta a$ half the minimal difference
between eigenvalues of $A$, which we will denote the strong limit
measurement, results in the standard von Neumann measurement. We will
show that when performing the strong limit measurement the probability
of measuring the probe in the domain $\left(a_{k}-\delta a,a_{k}+\delta a\right)$
is approximately $\left|\left\langle a_{k}|\psi_s\right\rangle \right|^{2}$
for $a_{k}$ an eigenvalue of $A$, and $\delta a$ the minimal difference
between eigenvalues of $A$. And the expectation of performing the
strong limit measurement is $g\mathbb{E}\left(A\right)$. The
variance is $g^{2}\text{Var}\left(A\right)+\Delta^{2}$.

\subsection{Strong limit for pure state \label{subsec:strong limit}}

We want to model strong measurements in a way we can compare them
easily to the weak value. For that we return to Eq. (\ref{eq:probe initial}).
Since $A$ is Hermitian, its spectral decomposition takes the form
$A=\sum_{i}a_{i}|a_{i}\rangle\langle a_{i}|$ for real $a_{i}$ and
$\left\{ |a_{i}\rangle\right\} $ an orthonormal basis and so $\sum_{i}|a_{i}\rangle\langle a_{i}|=I$,
and plugging it into Eq. (\ref{eq:before postselection}) we have 

\begin{align}
|\Phi\rangle & =e^{-ig A\otimes\mathcal{P}}|\psi_{s}\rangle\otimes|\phi\rangle\\
 & =\sum_{i}|a_{i}\rangle\left\langle a_{i}|e^{-ig\sum_{j}a_{j}|a_{j}\rangle\langle a_{j}|\otimes\mathcal{P}}|\psi_{s}\right\rangle \otimes|\phi\rangle\\
 & =\sum_{i}|a_{i}\rangle\left\langle a_{i}|\sum_{n}\frac{1}{n!}\left(-ig a_{i}\mathcal{P}\right)^{n}|\psi_{s}\right\rangle \otimes|\phi\rangle\\
 & =\sum_{i}|a_{i}\rangle\left\langle a_{i}|\psi_{s}\right\rangle e^{-ig a_{i}\mathcal{P}}|\phi\rangle\\
 & =\sum_{i}|a_{i}\rangle\left\langle a_{i}|\psi_{s}\right\rangle |\phi\left(q-g a_{i}\right)\rangle
\end{align}

and

\begin{align}
\left\langle q|\Phi\right\rangle  & =\sum_{i}|a_{i}\rangle\left\langle a_{i}|\psi_{s}\right\rangle \frac{1}{\left(2\pi\Delta^{2}\right)^{\frac{1}{4}}}\exp\left(-\frac{\left(q-g a_{i}\right)^{2}}{4\Delta^{2}}\right).\label{eq:psi q}
\end{align}

Hence,

\begin{align}
\int dq\left|\left\langle q|\Phi\right\rangle \right|^{2} & =\int dq\left|\sum_{i}|a_{i}\rangle\left\langle a_{i}|\psi_{s}\right\rangle \frac{1}{\left(2\pi\Delta^{2}\right)^{\frac{1}{4}}}\exp\left(-\frac{\left(q-g a_{i}\right)^{2}}{4\Delta^{2}}\right)\right|^{2}\\
 & =\frac{1}{\sqrt{2\pi\Delta^{2}}}\int dq\sum_{i}\left|\left\langle a_{i}|\psi_{s}\right\rangle \right|^{2}\exp\left(-\frac{\left(q-g a_{i}\right)^{2}}{2\Delta^{2}}\right).\label{eq:integrate psi q}
\end{align}

In the limit of strong measurement, we have small $\Delta$. Specifically
$\Delta\ll\gamma\delta a$ where $\delta a$ is half the minimal distance
between eigenvalues of $A$. And so the Gaussians do not overlap.
Now, if we measure the position of the probe, the probability of it
being in $\left[a_{k}-\delta a,\,\,a_{k}+\delta a\right]$ is given
by

\begin{align}
\int\limits _{a_{k}-\delta a}^{a_{k}+\delta a}dq\left|\left\langle q|\Phi\right\rangle \right|^{2} & =\frac{1}{\sqrt{2\pi\Delta^{2}}}\int\limits _{a_{k}-\delta a}^{a_{k}+\delta a}dq\sum_{i}\left|\left\langle a_{i}|\psi_{s}\right\rangle \right|^{2}\exp\left(-\frac{\left(q-g a_{i}\right)^{2}}{2\Delta^{2}}\right)\simeq\left|\left\langle a_{k}|\psi_{s}\right\rangle \right|^{2}.
\end{align}

Similarly to Eq. (\ref{eq:integrate psi q}), and since $|\Phi\rangle$
is normalized we have 

\begin{align}
\mathbb{E}\left(Q\right) & =\int dq\cdot q\left|\left\langle q|\Phi\right\rangle \right|^{2}\nonumber \\
 & =\frac{1}{\sqrt{2\pi\Delta^{2}}}\sum_{i}\left|\left\langle a_{i}|\psi_{s}\right\rangle \right|^{2}\sqrt{2\pi\Delta^{2}}g a_{i}\\
 & =g\left\langle \psi_{s}|A|\psi_{s}\right\rangle 
\end{align}

and

\begin{align}
\left\langle \Phi|Q^{2}|\Phi\right\rangle  & =\int dq\cdot q^{2}\left|\left\langle q|\Phi\right\rangle \right|^{2}\\
 & =\frac{1}{\sqrt{2\pi\Delta^{2}}}\sum_{i}\left|\left\langle a_{i}|\psi_{s}\right\rangle \right|^{2}\sqrt{2\pi\Delta^{2}}\left(g^{2}a_{i}^{2}+\Delta^{2}\right)\\
 & =g^{2}\left\langle \psi_{s}|A^{2}|\psi_{s}\right\rangle +\Delta^{2}.
\end{align}

Therefore,
\begin{equation}
\text{Var}\left(Q\right)_{\Phi}=\left\langle \Phi|Q^{2}|\Phi\right\rangle -\left\langle \Phi|Q|\Phi\right\rangle ^{2}=g^{2}\text{Var}\left(A\right)_{\psi_{s}}+\Delta^{2}.
\end{equation}

\subsection{Strong limit for mixed state\label{subsec:strong mixed}}

The initial state of the system is $\rho_s$, while the initial state
of the probe and the interaction Hamiltonian are the same as above.
So after this interaction, the joint system state is

\begin{equation}
\rho_{s,p}=\exp\left(-ig A\otimes\mathcal{P}\right)\left(\rho_{s}\otimes|\phi\rangle\langle\phi|\right)\exp\left(ig A\otimes\mathcal{P}\right).
\end{equation}

We will again look at the spectral decomposition of the operator $A$,
$A=\sum_{i}a_{i}|a_{i}\rangle\langle a_{i}|$ and plug in $\sum_{i}|a_{i}\rangle\langle a_{i}|=I$,
yielding

\begin{align}
\rho_{s,p} & =\sum_{i,j}|a_{i}\rangle\left\langle a_{i}|\exp\left(-ig\sum_{k}a_{k}|a_{k}\rangle\langle a_{k}|\otimes\mathcal{P}\right)\left(\rho_{s}\otimes|\phi\rangle\langle\phi|\right)\exp\left(ig\sum_{l}a_{l}|a_{l}\rangle\langle a_{l}|\otimes\mathcal{P}\right)|a_{j}\right\rangle \langle a_{j}|\\
 & =\sum_{i,j}|a_{i}\rangle\left\langle a_{i}|\exp\left(-ig a_{i}\mathcal{P}\right)\left(\rho_{s}\otimes|\phi\rangle\langle\phi|\right)\exp\left(ig a_{j}\mathcal{P}\right)|a_{j}\right\rangle \langle a_{j}|\\
 & =\sum_{i,j}\left\langle a_{i}|\rho_{s}|a_{j}\right\rangle |a_{i}\rangle\langle a_{j}|\otimes\exp\left(-ig a_{i}\mathcal{P}\right)|\phi\rangle\langle\phi|\exp\left(ig a_{j}\mathcal{P}\right)\\
 & =\iint dqdq'\sum_{i,j}\left\langle a_{i}|\rho_{s}|a_{j}\right\rangle |a_{i}\rangle\langle a_{j}|\otimes\phi\left(q-g a_{i}\right)|q\rangle\langle q'|\phi^{*}\left(q'-g a_{j}\right),
 \label{eq:joint state for strong}
\end{align}

which is a weighted mixed-state sum of shifted Gaussians. From now
on we want to use only the probe and disregard the system, so we will
trace it out:

\begin{align}
\rho_{p} & =\sum_{k}\left\langle a_{k}|\iint dqdq'\sum_{i,j}\left\langle a_{i}|\rho_{s}|a_{j}\right\rangle |a_{i}\rangle\langle a_{j}|\otimes\phi\left(q-g a_{i}\right)|q\rangle\langle q'|\phi^{*}\left(q'-g a_{j}\right)|a_{k}\right\rangle \\
 & =\iint dqdq'\sum_{k}\left\langle a_{k}|\rho_{s}|a_{k}\right\rangle \phi\left(q-g a_{k}\right)|q\rangle\langle q'|\phi^{*}\left(q'-g a_{k}\right).
\end{align}

Notice that $\rho_{p}$ is normalized, and $Pr\left(x\right)=\text{Tr}\left(\Pi_{x}\rho\right)$
for $\Pi_{x}=|x\rangle\langle x|$ and normalized $\rho$ and so 
\begin{align}
Pr\left(x\right) & =\iint dqdq'\sum_{k}\left\langle a_{k}|\rho_{s}|a_{k}\right\rangle \phi\left(q-g a_{k}\right)\left\langle x|q\right\rangle \left\langle q'|x\right\rangle \phi^{*}\left(q'-g a_{k}\right)\\
 & =\sum_{k}\left\langle a_{k}|\rho_{s}|a_{k}\right\rangle \left|\phi\left(x-g a_{k}\right)\right|^{2}.
\end{align}

Hence, similarly to the pure case, we have $Pr\left(x=g a_{i}\right)\simeq\left\langle a_{i}|\rho_{s}|a_{i}\right\rangle $
while if $x\not\simeq g a_{i}$ for any $i$ all terms in the
sum will be approximately zero and so 
\begin{equation}
Pr\left(x\right)\simeq\begin{cases}
\left\langle a_{i}|\rho_{s}|a_{i}\right\rangle  & x=g a_{i}\\
0 & x\neq g a_{j}\text{ for any }j
\end{cases}.
\end{equation}
And this is indeed a strong measurement.

\subsection{Strong measurement with postselection \label{derivation strong measurement with post selection}}

We have the joint state in Eq. (\ref{eq:joint state for strong}), but this time,
instead of tracing out the system, we post-select on it being in state
$|\psi_{f}\rangle$, just like we did for the weak-value setup. In
this case we obtain the un-normalized state
\begin{equation}
\rho_{p}\simeq\iint dqdq'\sum_{i,j}\left\langle a_{i}|\rho|a_{j}\right\rangle \left\langle \psi_{f}|a_{i}\right\rangle \left\langle a_{j}|\psi_{f}\right\rangle \phi\left(q-\gamma a_{i}\right)|q\rangle\langle q'|\phi^{*}\left(q'-\gamma a_{j}\right)
\end{equation}
Now, to obtain a normalized state we will need to divide by the probability to obtain the desired postselection, i.e. $\langle \psi_f | \rho | \psi_f \rangle$, but since we will only be interested in the cases where the linear order vanishes, then the denominator will be of no interest to us, and we will continue with the un-normalized state.
Now, the expectation value of the position of the probe is given (up to normalization) by
\begin{align}
\left\langle Q\right\rangle _{\rho_{p}} & \propto \int dq''\left\langle q''|\rho_{p}|q''\right\rangle q''\\
 & \simeq\iiint dqdq'dq''q''\sum_{i,j}\left\langle a_{i}|\rho|a_{j}\right\rangle \left\langle \psi_{f}|a_{i}\right\rangle \left\langle a_{j}|\psi_{f}\right\rangle \phi\left(q-\gamma a_{i}\right)\left\langle q''|q\right\rangle \left\langle q'|q''\right\rangle \phi^{*}\left(q'-\gamma a_{j}\right)\\
 & \simeq\int dq''q''\sum_{i,j}\left\langle a_{i}|\rho|a_{j}\right\rangle \left\langle \psi_{f}|a_{i}\right\rangle \left\langle a_{j}|\psi_{f}\right\rangle \phi\left(q''-\gamma a_{i}\right)\phi^{*}\left(q''-\gamma a_{j}\right)
\end{align}

Now, since each of these $\phi\left(q''-\gamma a_{i}\right)$ are peaked
around $a_{i}$, they will be non-zero only where $a_{i}=a_{j}$,
in which case

\begin{align}
\left\langle Q\right\rangle _{\rho_{p}} & \simeq\int dq''q''\sum_{i}\left\langle a_{i}|\rho|a_{i}\right\rangle \left\langle \psi_{f}|a_{i}\right\rangle \left\langle a_{i}|\psi_{f}\right\rangle \phi\left(q''-\gamma a_{i}\right)\phi^{*}\left(q''-\gamma a_{i}\right)\\
 & \simeq\sum_{i}\gamma a_{i}\left\langle \psi_{f}|a_{i}\right\rangle \left\langle a_{i}|\rho|a_{i}\right\rangle \left\langle a_{i}|\psi_{f}\right\rangle 
\end{align}
\begin{equation}
\left\langle Q\right\rangle _{strong\,with\,post}\simeq\gamma\sum_{i}a_{i}\left\langle \psi_{f}|a_{i}\right\rangle \left\langle a_{i}|\rho|a_{i}\right\rangle \left\langle a_{i}|\psi_{f}\right\rangle \label{eq:strong with post}
\end{equation}

\subsubsection{Coinciding with PVM with postselection}

We start with an initial state $\rho$ and then measure the the operator
$A$ on it, For $A=a_{i}|a_{i}\rangle\langle a_{i}|$.
If the measurement result was $a_{i}$ then the state collapses to
$|a_{i}\rangle$. This happens with probability $\left\langle a_{i}|\rho|a_{i}\right\rangle $,
and so the state after the measurement, if the result measurement
is still unknown, is given by the classical super-position over eigen-states:
\begin{equation}
\rho_{a.m.}=\sum_{i}\left\langle a_{i}|\rho|a_{i}\right\rangle |a_{i}\rangle\langle a_{i}|
\end{equation}

Next, the probability to now measure the final state $|\psi_f\rangle$ is given by the projection of the state on $|\psi_f\rangle$:
\begin{equation}
Pr\left(|\psi_f\rangle\right)=\sum_{i}\left\langle a_{i}|\rho|a_{i}\right\rangle \left\langle \psi_{f}|a_{i}\right\rangle \left\langle a_{i}|\psi_{f}\right\rangle 
\end{equation}
and so the joint probability of $|a_i\rangle$ and $|\psi_f\rangle$ is 
\begin{equation}
Pr\left(a_{i}, |\psi_f\rangle \right)=\left\langle a_{i}|\rho|a_{i}\right\rangle \left\langle \psi_{f}|a_{i}\right\rangle \left\langle a_{i}|\psi_{f}\right\rangle 
\end{equation}
In order to obtain the probability for $|a_i\rangle$ when we postselect on $|\psi_f\rangle$ we now need to divide by the probability of indeed measuring $|\psi_f\rangle$, which is $\langle \psi_f|\rho|\psi_f\rangle$ (since the probability of measuring $|\psi_f\rangle$ does not change when we measure in the basis of $\{a_i\}$ without learning the result), leading to
\begin{equation}
Pr\left(a_{i} \mid  |\psi_f\rangle\right)= \frac{1}{\langle \psi_f | \rho | \psi_f \rangle}
\left\langle a_{i}|\rho|a_{i}\right\rangle \left\langle \psi_{f}|a_{i}\right\rangle \left\langle a_{i}|\psi_{f}\right\rangle. 
\label{eq:prob_eigenvalue}
\end{equation}

And so this is the probability of obtaining measurement result $a_{i}$
when starting at state $\rho$ and post-selecting on $|\psi_{f}\rangle$.
And so the expectation of the measurement of $A$, with the post selection
is 
\begin{equation}
\sum_{i}Pr\left(a_{i} \mid  |\psi_f\rangle\right)a_{i}=\frac{1}{\langle \psi_f | \rho | \psi_f \rangle} \sum_{i}a_{i}\left\langle a_{i}|\rho|a_{i}\right\rangle \left\langle \psi_{f}|a_{i}\right\rangle \left\langle a_{i}|\psi_{f}\right\rangle 
\label{eq:expectation strong post}
\end{equation}
which is the same as in Eq. (\ref{eq:strong with post}) up to the normalization factor.
It is important to note that the possible results of the measurement are the eigenvalues of the operator 
\cite{greensite2003lecture}, and the probabilities of each eigenvalue is given by Eq. (\ref{eq:prob_eigenvalue}) and so the expectation value holds information about both eigenvalues and eigenvectors. And so with clever choices of initial and final states we may be able to fully learn the operator A through the expectation value in Eq. (\ref{eq:expectation strong post}). Note that the transition from Eq. (\ref{eq:prob_eigenvalue}) to Eq. (\ref{eq:expectation strong post}) is because the measurement results are the eigenvalues, and it holds whether or not we have prior knowledge regarding the values of the eigenvalues.

\section{Lindblad Noise Approximated by Noise Before Ideal WMP \label{app: lindblad approximation}}

The full initial system consists of the primary system and the probe,
and is given by $\rho=\rho_{\text{primay}}\otimes\rho_{\text{probe}}$.
The propagation of the full system is determined by the Master equation
in Lindblad form

\begin{equation}
\frac{\partial\rho}{\partial t}=\mathcal{L}\left(\rho\right)=-i\left[H,\rho\right]+\mathcal{D}\left[\rho\right],
\end{equation}
\begin{equation}
\mathcal{D}\left[\rho\right]=\sum_{k}\gamma_{k}\left[L_{k}\rho L_{k}^{\dagger}-\frac{1}{2}\left\{ L_{k}^{\dagger}L_{k},\rho\right\} \right],
\end{equation}

where $H$ is the Hamiltonian governing the weak interaction and $\mathcal{D}$
is the dissipator which gives rise to the noise. For simplicity we
will assume that the weak interaction Hamiltonian is constant over
time, and for this section we will define the weak value parameter $g=\tilde{g}t$. Under this definition the Hamiltonian is

\begin{equation}
H=\tilde{g}A\otimes \mathcal{P}.
\end{equation}

Converting to the Choi representation we can replace the density matrix
$\rho=\sum_{i,j}\rho_{ij}|i\rangle\langle j|$ by the vector
\begin{equation}
\text{vec}\left(\rho\right)=\sum_{i,j}\rho_{ij}|j\rangle\otimes|i\rangle.
\end{equation}

In this representation, the Master equation becomes
$
\frac{d}{dt}\text{vec}\left(\rho\right)=\hat{\mathcal{L}}\text{vec}\left(\rho\right)
$
where $\hat{\mathcal{L}}$ is a superoperator given by
\begin{equation}
\hat{\mathcal{L}}=-i\left(I\otimes H-H^{T}\otimes I\right)+\sum_{k}\gamma_{k}\left[L_{k}^{*}\otimes L_{k}-\frac{1}{2}I\otimes L_{k}^{\dagger}L_{k}-\frac{1}{2}\left(L_{k}L_{k}^{\dagger}\right)^{T}\otimes I\right],\label{eq:lindblad for vector}
\end{equation}

for which the solution is 
\begin{equation}
\text{vec}\left(\rho\left(t\right)\right)=e^{\hat{\mathcal{L}}t}\text{vec}\left(\rho\left(0\right)\right).
\end{equation}
Let us define 
$\tilde{\gamma}=\max_{k}\gamma_{k}$
and
$\lambda_{k}=\frac{\gamma_{k}}{\tilde{\gamma}}$
 and so $\lambda_{k}<1$. Plugging these definitions into Eq. (\ref{eq:lindblad for vector})
we have
\begin{equation}
\hat{\mathcal{L}}=-i\left(I\otimes H-H^{T}\otimes I\right)+\tilde{\gamma}\sum_{k}\lambda_{k}\left[L_{k}^{*}\otimes L_{k}-\frac{1}{2}I\otimes L_{k}^{\dagger}L_{k}-\frac{1}{2}\left(L_{k}L_{k}^{\dagger}\right)^{T}\otimes I\right].
\end{equation}

If we now separate 
\begin{equation}
\hat{\mathcal{L}}=\tilde{g}\hat{\mathcal{L}}_{H}+\tilde{\gamma}\hat{\mathcal{L}}_{L}
\end{equation}
for
\begin{equation}
\hat{\mathcal{L}}_{H}=\frac{-i\left(I\otimes H-H^{T}\otimes I\right)}{\tilde{g}}=-i\left(I\otimes\left(A_{\text{primary}}\otimes P_{\text{probe}}\right)-\left(A_{\text{primary}}\otimes P_{\text{probe}}\right)^{T}\otimes I\right)
\end{equation}
and
\begin{equation}
\hat{\mathcal{L}}_{L}=\sum_{k}\lambda_{k}\left[L_{k}^{*}\otimes L_{k}-\frac{1}{2}I\otimes L_{k}^{\dagger}L_{k}-\frac{1}{2}\left(L_{k}L_{k}^{\dagger}\right)^{T}\otimes I\right],
\end{equation}

we have 
\begin{equation}
\text{vec}\left(\rho\left(t\right)\right)=e^{\tilde{g}\hat{\mathcal{L}}_{H}t+\tilde{\gamma}\hat{\mathcal{L}}_{L}t}\text{vec}\left(\rho\left(0\right)\right).
\end{equation}

We can now expand in Taylor series to obtain
\begin{align}
e^{\tilde{g}t\hat{\mathcal{L}}_{H}+\tilde{\gamma}t\hat{\mathcal{L}}_{L}} & =I+\tilde{g}\hat{\mathcal{L}}_{H}t+\tilde{\gamma}\hat{\mathcal{L}}_{L}t+O\left(\tilde{g}t\tilde{\gamma}t,\left(g_{WV}t\right)^{2},\left(\tilde{\gamma}t\right)^{2}\right)\\
 & =\left(I+\tilde{g}\hat{\mathcal{L}}_{H}t\right)\left(I+\tilde{\gamma}\hat{\mathcal{L}}_{L}t\right)+O\left(\tilde{g}t\tilde{\gamma}t,\left(\tilde{g}t\right)^{2},\left(\tilde{\gamma}t\right)^{2}\right)\\
 & =e^{\tilde{g}t\hat{\mathcal{L}}_{H}}e^{\tilde{\gamma}t\hat{\mathcal{L}}_{L}}+O\left(\tilde{g}t\tilde{\gamma}t,\left(\tilde{g}t\right)^{2},\left(\tilde{\gamma}t\right)^{2}\right).
\end{align}

Now, $e^{\tilde{\gamma}t\hat{\mathcal{L}}_{L}}$ is the propagation due to
a noise channel and $e^{\tilde{g}t\hat{\mathcal{L}}_{H}}$ is the propagation
due to the weak interaction. And so we showed that a noise channel
followed by a noise-less weak interaction is a good approximation
of a Master equation consisting of the intentional weak interaction
and the accompanying noise terms.

To specify the requirements for the approximation validity more clearly,
let us expand to a higher order. For brevity of notation we will denote
$O\left((\tilde{\cdot}t)^3\right)=O\left(\left(\tilde{g}t\right)^{2}\tilde{\gamma}t,\left(\tilde{\gamma}t\right)^{2}\tilde{g}t,\left(\tilde{g}t\right)^{3},\left(\tilde{\gamma}t\right)^{3}\right)$. We obtain

\begin{align}
 & e^{\tilde{g}t\hat{\mathcal{L}}_{H}+\tilde{\gamma}t\hat{\mathcal{L}}_{L}}\\
 & =I+\tilde{g}\hat{\mathcal{L}}_{H}t+\tilde{\gamma}\hat{\mathcal{L}}_{L}t+\frac{1}{2}\left(\tilde{g}t\hat{\mathcal{L}}_{H}+\tilde{\gamma}t\hat{\mathcal{L}}_{L}\right)^{2}+O\left((\tilde{\cdot}t)^3\right)\\
 & =I+\tilde{g}\hat{\mathcal{L}}_{H}t+\tilde{\gamma}\hat{\mathcal{L}}_{L}t+\frac{1}{2}\left(\tilde{g}t\right)^{2}\hat{\mathcal{L}}_{H}^{2}+\frac{1}{2}\left(\tilde{\gamma}t\right)^{2}\hat{\mathcal{L}}_{L}^{2}+\frac{1}{2}\tilde{g}t\tilde{\gamma}t\left\{ \hat{\mathcal{L}}_{H},\hat{\mathcal{L}}_{L}\right\} +O\left((\tilde{\cdot}t)^3\right).
\end{align}
On the other hand
\begin{align}
 & e^{\tilde{g}t\hat{\mathcal{L}}_{H}}e^{\tilde{\gamma}t\hat{\mathcal{L}}_{L}}\\
 & =\left(I+\tilde{g}t\hat{\mathcal{L}}_{H}+\frac{1}{2}\left(\tilde{g}t\right)^{2}\hat{\mathcal{L}}_{H}^{2}\right)\left(I+\tilde{\gamma}t\hat{\mathcal{L}}_{L}+\frac{1}{2}\left(\tilde{\gamma}t\right)^{2}\hat{\mathcal{L}}_{L}^{2}\right)+O\left((\tilde{\cdot}t)^3\right)\\
 & =I+\tilde{\gamma}\hat{\mathcal{L}}_{L}t+\frac{1}{2}\left(\tilde{\gamma}t\right)^{2}\hat{\mathcal{L}}_{L}^{2}+\tilde{g}t\hat{\mathcal{L}}_{H}\left(I+\tilde{\gamma}t\hat{\mathcal{L}}_{L}\right)+\frac{1}{2}\left(\tilde{g}t\right)^{2}\hat{\mathcal{L}}_{H}^{2}+O\left((\tilde{\cdot}t)^3\right)\\
 & =I+\tilde{\gamma}\hat{\mathcal{L}}_{L}t+\frac{1}{2}\left(\tilde{\gamma}t\right)^{2}\hat{\mathcal{L}}_{L}^{2}+\tilde{g}t\hat{\mathcal{L}}_{H}+\tilde{g}t\tilde{\gamma}t\hat{\mathcal{L}}_{H}\hat{\mathcal{L}}_{L}+\frac{1}{2}\left(\tilde{g}t\right)^{2}\hat{\mathcal{L}}_{H}^{2}+O\left((\tilde{\cdot}t)^3\right)\\
 & =I+\tilde{g}t\hat{\mathcal{L}}_{H}+\tilde{\gamma}\hat{\mathcal{L}}_{L}t+\frac{1}{2}\left(\tilde{g}t\right)^{2}\hat{\mathcal{L}}_{H}^{2}+\frac{1}{2}\left(\tilde{\gamma}t\right)^{2}\hat{\mathcal{L}}_{L}^{2}+\frac{1}{2}\tilde{g}t\tilde{\gamma}t\left\{ \hat{\mathcal{L}}_{H},\hat{\mathcal{L}}_{L}\right\} -\frac{1}{2}\tilde{g}t\tilde{\gamma}t\left[\hat{\mathcal{L}}_{L},\hat{\mathcal{L}}_{H}\right]+O\left((\tilde{\cdot}t)^3\right).
\end{align}

And so 
\begin{equation}
e^{\tilde{g}t\hat{\mathcal{L}}_{H}+\tilde{\gamma}t\hat{\mathcal{L}}_{L}}=e^{\tilde{g}t\hat{\mathcal{L}}_{H}}e^{\tilde{\gamma}t\hat{\mathcal{L}}_{L}}-\frac{1}{2}\tilde{g}t\tilde{\gamma}t\left[\hat{\mathcal{L}}_{L},\hat{\mathcal{L}}_{H}\right]+O\left(\left(\tilde{g}t\right)^{2}\tilde{\gamma}t,\left(\tilde{\gamma}t\right)^{2}\tilde{g}t,\left(\tilde{g}t\right)^{3},\left(\tilde{\gamma}t\right)^{3}\right).
\label{eq:noise during as noise before}
\end{equation}

And so approximating the noise as a noise channel acting before the entanglement, which is done by disregarding all terms apart from the first term in the RHS of Eq. (\ref{eq:noise during as noise before}), is valid when
$\frac{1}{2}\tilde{g}t\tilde{\gamma}t\left|\left|\left[\hat{\mathcal{L}}_{L},\hat{\mathcal{L}}_{H}\right]\right|\right|\ll \tilde{\gamma}t\left|\left|\mathcal{\hat{L}}_{L}\right|\right|$
and $\frac{1}{2}\tilde{g}t\tilde{\gamma}t\left|\left|\left[\hat{\mathcal{L}}_{L},\hat{\mathcal{L}}_{H}\right]\right|\right|\ll \tilde{g}t\left|\left|\mathcal{\hat{L}}_{H}\right|\right|$,
which can be presented as the conditions:
\begin{equation}
\tilde{g}t\ll2\frac{\left|\left|\mathcal{\hat{L}}_{L}\right|\right|}{\left|\left|\left[\hat{\mathcal{L}}_{L},\hat{\mathcal{L}}_{H}\right]\right|\right|}
\end{equation}
and
\begin{equation}
\tilde{\gamma}t\ll2\frac{\left|\left|\mathcal{\hat{L}}_{H}\right|\right|}{\left|\left|\left[\hat{\mathcal{L}}_{L},\hat{\mathcal{L}}_{H}\right]\right|\right|}.
\end{equation}

This will hold for most physical cases where $\gamma=\tilde{\gamma}t$ and
$g=\tilde{g}t$ are small.

\section{Proof of Theorem - Weak Value advantage at learning $A$ under Amplitude and Phase damping \label{app:main theormen proof}}

\subsection{The combined channel }

We will now show that for two noise channels $\mathcal{E}_{2},\mathcal{E}_{1}$
with noise parameters $p_{i}=\lambda_{i}\gamma$ then $\gamma$ is
a noise parameter for the combined channel $\mathcal{E}_{2}\circ\mathcal{E}_{1}$.
We denote the WV when the noise channel is $\mathcal{E}$ by $A_w\left(\mathcal{E}\right)$.
Then the weak value of the combined channel is

\begin{equation}
A_{w}\left(\mathcal{E}_{2}\circ\mathcal{E}_{1}\right)=\frac{\left\langle \psi_{f}|A\mathcal{E}_{2}\left(\mathcal{E}_{1}\left(\rho\right)\right)|\psi_{f}\right\rangle }{\left\langle \psi_{f}|\mathcal{E}_{2}\left(\mathcal{E}_{1}\left(\rho\right)\right)|\psi_{f}\right\rangle }=A_{w}\left(\mathcal{E}_{1}\right)+p_{2}\Delta_{\mathcal{E}_{2}}\left(\mathcal{E}_{1}\right)=A_{w}\left(\mathcal{E}_{1}\right)+\lambda_{2}\gamma\Delta_{\mathcal{E}_{2}}\left(\mathcal{E}_{1}\right),
\end{equation}
where 
$\Delta_{\mathcal{E}_{2}}\left(\mathcal{E}_{1}\right)$ refers to the linear order affect of the noise channel $\mathcal{E}_{2}$, while the initial state is assumed to be $\mathcal{E}_{1}$, instead of $\rho$. In other words, when the Taylor expansion is done only for the noise channel $\mathcal{E}_{2}$, where the noise channel $\mathcal{E}_{1}$ is not yet dealt with.
$A_{w}\left(\mathcal{E}_{1}\right)$ and $\Delta_{\mathcal{E}_{2}}\left(\mathcal{E}_{1}\right)$
And so there
still remains a dependency on $\mathcal{E}_{1}$ inside these terms.
Now,
\begin{equation}
A_{w}\left(\mathcal{E}_{1}\right)=A_{w}+\gamma\lambda_{1}\Delta_{\mathcal{E}_{1}}.
\end{equation}

Since we only want the linear order in $\gamma$, we will take only
the zeroth order of $\gamma$ in $\Delta_{\mathcal{E}_{2}}\left(\mathcal{E}_{1}\right)$,
for which by definition $\Delta_{\mathcal{E}_{2}}\left(\mathcal{E}_{1}\right)=\Delta_{\mathcal{E}_{2}}$,
and so to linear order in $\gamma$ we have 
\begin{equation}
A_{w}\left(\mathcal{E}_{2}\circ\mathcal{E}_{1}\right)=A_{w}+\gamma\lambda_{1}\Delta_{\mathcal{E}_{1}}+\gamma\lambda_{2}\Delta_{\mathcal{E}_{2}}.
\end{equation}

Hence, whenever the linear order vanishes for any of the separate
channels it will also vanish for the combined channel.

\subsection{Claim 1 -- WMP can accomplish the task}

The two separate channels we are interested in are:
\begin{itemize}
\item Amplitude damping: $E_{0}=\left(\begin{matrix}1 & 0\\
0 & \sqrt{1-\gamma}
\end{matrix}\right)$, $E_{1}=\left(\begin{matrix}0 & \sqrt{\gamma}\\
0 & 0
\end{matrix}\right)$
\item Phase damping: $E_{0}=\left(\begin{matrix}1 & 0\\
0 & \sqrt{1-\lambda}
\end{matrix}\right)$, $E_{1}=\left(\begin{matrix}0 & 0\\
0 & \sqrt{\lambda}
\end{matrix}\right)$. 
\end{itemize}
For amplitude damping we have

\begin{equation}
E_{0}\rho E_{0}^{\dagger}+E_{1}\rho E_{1}^{\dagger}=\left(\begin{matrix}\rho_{11}+\gamma\rho_{22} & \sqrt{1-\gamma}\rho_{12}\\
\sqrt{1-\gamma}\rho_{21} & \left(1-\gamma\right)\rho_{22}
\end{matrix}\right).
\end{equation}

Now, to first order in $g$ we have 
\begin{equation}
\mathcal{E}_{\text{a.d.}}\left(\rho\right)=E_{0}\rho E_{0}^{\dagger}+E_{1}\rho E_{1}^{\dagger}\simeq\rho+\gamma M^{\text{ad}},
\end{equation}
where we define $M^{\text{ad}}=\left(\begin{matrix}\rho_{22} & -\frac{1}{2}\rho_{12}\\
-\frac{1}{2}\rho_{21} & -\rho_{22}
\end{matrix}\right)$. And so, keeping only leading order in $g$: 
\begin{align}
A_{w,\mathcal{E}_{a.d}} & =\frac{\left\langle \psi_{f}|A\left(\rho+g M^{\text{ad}}\right)|\psi_{f}\right\rangle }{\left\langle \psi_{f}|\rho+g M^{\text{ad}}|\psi_{f}\right\rangle }\\
 & =\frac{\left\langle \psi_{f}|A\rho|\psi_{f}\right\rangle +g\left\langle \psi_{f}|AM^{\text{ad}}|\psi_{f}\right\rangle }{\left\langle \psi_{f}|\rho|\psi_{f}\right\rangle \left(1+g\frac{\left\langle \psi_{f}|M^{\text{ad}}|\psi_{f}\right\rangle }{\left\langle \psi_{f}|\rho|\psi_{f}\right\rangle }\right)}\\
 & \simeq\frac{\left\langle \psi_{f}|A\rho|\psi_{f}\right\rangle }{\left\langle \psi_{f}|\rho|\psi_{f}\right\rangle }+g\left(\frac{\left\langle \psi_{f}|AM^{\text{ad}}|\psi_{f}\right\rangle }{\left\langle \psi_{f}|\rho|\psi_{f}\right\rangle }-\frac{\left\langle \psi_{f}|A\rho|\psi_{f}\right\rangle }{\left\langle \psi_{f}|\rho|\psi_{f}\right\rangle }\frac{\left\langle \psi_{f}|M^{\text{ad}}|\psi_{f}\right\rangle }{\left\langle \psi_{f}|\rho|\psi_{f}\right\rangle }\right).
\end{align}

And so 
\[
\Delta_{\text{\text{amplitude damping}}}=\frac{\left\langle f|AM^{\text{ad}}|f\right\rangle }{\left\langle f|\rho|f\right\rangle }-\frac{\left\langle f|M^{\text{ad}}|f\right\rangle }{\left\langle f|\rho|f\right\rangle }\frac{\left\langle f|A\rho|f\right\rangle }{\left\langle f|\rho|f\right\rangle }.
\]

Phase damping is equivalent to phase flip, which is a Pauli channel with the pauli $Z$ and so 

\begin{equation}
    \Delta_{\text{\text{phase damping}}}=\frac{\left\langle \psi_{f}|AZ\rho_{s}Z|\psi_{f}\right\rangle }{\left\langle \psi_{f}|\rho_{s}|\psi_{f}\right\rangle }-\frac{\left\langle \psi_{f}|Z\rho_{s}Z|\psi_{f}\right\rangle }{\left\langle \psi_{f}|\rho_{s}|\psi_{f}\right\rangle }\frac{\left\langle \psi_{f}|A\rho_{s}|\psi_{f}\right\rangle }{\left\langle \psi_{f}|\rho_{s}|\psi_{f}\right\rangle }.
\end{equation}

Next, we solved the euqations $\Delta_{\text{\text{amplitude damping}}}=0$ and $\Delta_{\text{\text{phase damping}}}=0$ and found that
in order to overcome amplitude damping and phase damping simultaneously the initial and final states must be:

\begin{equation}
|\psi_{f}\rangle=\left(\begin{matrix}1\\
0
\end{matrix}\right),\rho=\left(\begin{matrix}\rho_{11} & 0\\
0 & 1-\rho_{11}
\end{matrix}\right),\left\langle \psi_{f}|\rho|\psi_{f}\right\rangle =\rho_{11}\neq 0,A_{w}=a_{11},
\end{equation}
\begin{equation}
|\psi_{f}\rangle=\left(\begin{matrix}0\\
1
\end{matrix}\right),\rho=\left(\begin{matrix}\rho_{11} & 0\\
0 & 1-\rho_{11}
\end{matrix}\right),\left\langle \psi_{f}|\rho|\psi_{f}\right\rangle =1-\rho_{11}\neq 0,A_{w}=a_{22},
\end{equation}
or
\begin{equation}
|\psi_{f}\rangle=\left(\begin{matrix}f_{1}\\
f_{2}
\end{matrix}\right),\rho=\left(\begin{matrix}1 & 0\\
0 & 0
\end{matrix}\right),\left\langle \psi_{f}|\rho|\psi_{f}\right\rangle =\left|f_{1}\right|^{2}\neq 0,A_{w}=a_{11}+a_{21}\frac{f_{2}^{*}}{f_{1}^{*}}.
\end{equation}

Notice that the first two sets of states can overcome Pauli noise as well, as presented above.
Now, learning $A$ using this. We get $a_{11}$ and $a_{22}$ from
the first two cases. Now, for the third case, if we chose $f_{1}=f_{2}=\frac{1}{\sqrt{2}}$
we get 
$A_{w}=a_{11}+a_{21}$
And since the shift is the real part of the weak value we can use
this to learn $\mathcal{R}\left(a_{12}\right)$. And to learn $\mathcal{I}\left(a_{12}\right)$
we use $f_{1}=\frac{1}{\sqrt{2}}$ and $f_{2}=\frac{i}{\sqrt{2}}$
and so 
$A_{w}=a_{11}-ia_{21}$.

\subsection{Claim 2 -- strong measurements cannot accomplish the task even for amplitude damping alone}

For strong measurements 
$\left\langle Q\right\rangle =\gamma \text{Tr} \left(A\mathcal{E}\left(\rho\right)\right)$.

\begin{equation}
\text{Tr}\left(A\rho\right)=a_{11}\rho_{11}+a_{12}\rho_{21}+a_{21}\rho_{12}+a_{22}\rho_{22}
\end{equation}

For amplitude damping we have
\begin{align}
\mathcal{E}_{AD}\left(\rho\right) & =\left(\begin{matrix}1 & 0\\
0 & \sqrt{1-\gamma}
\end{matrix}\right)\left(\begin{matrix}\rho_{11} & \rho_{12}\\
\rho_{21} & \rho_{22}
\end{matrix}\right)\left(\begin{matrix}1 & 0\\
0 & \sqrt{1-\gamma}
\end{matrix}\right)+\left(\begin{matrix}0 & \sqrt{\gamma}\\
0 & 0
\end{matrix}\right)\left(\begin{matrix}\rho_{11} & \rho_{12}\\
\rho_{21} & \rho_{22}
\end{matrix}\right)\left(\begin{matrix}0 & 0\\
\sqrt{\gamma} & 0
\end{matrix}\right)\\
 &  =\left(\begin{matrix}\rho_{11}+\gamma\rho_{22} & \sqrt{1-\gamma}\rho_{12}\\
\sqrt{1-\gamma}\rho_{21} & \left(1-\gamma\right)\rho_{22}
\end{matrix}\right)
\end{align}

\begin{align}
\text{Tr}\left(A\mathcal{E}_{AD}\left(\rho\right)\right) & =\text{Tr}\left(\left(\begin{matrix}a_{11} & a_{12}\\
a_{21} & a_{22}
\end{matrix}\right)\left(\begin{matrix}\rho_{11}+\gamma\rho_{22} & \sqrt{1-\gamma}\rho_{12}\\
\sqrt{1-\gamma}\rho_{21} & \left(1-\gamma\right)\rho_{22}
\end{matrix}\right)\right)\\
 & =a_{11}\left(\rho_{11}+\gamma\rho_{22}\right)+a_{12}\sqrt{1-\gamma}\rho_{21}+a_{21}\sqrt{1-\gamma}\rho_{12}+a_{22}\left(1-\gamma\right)\rho_{22}
\end{align}

And so 
\begin{align}
\text{Tr}\left(A\mathcal{E}_{AD}\left(\rho\right)\right)-\text{Tr}\left(A\rho\right) & =a_{11}\gamma\rho_{22}+a_{12}\left(\sqrt{1-\gamma}-1\right)\rho_{21}+a_{21}\left(\sqrt{1-\gamma}-1\right)\rho_{12}-a_{22}\gamma\rho_{22}\\
 & =a_{11}\gamma\rho_{22}+a_{12}\left(-\frac{1}{2}\gamma+O\left(\gamma^{2}\right)\right)\rho_{21}+a_{21}\left(-\frac{1}{2}\gamma+O\left(\gamma^{2}\right)\right)\rho_{12}-a_{22}\gamma\rho_{22}
\end{align}

And so for the linear order of $\gamma$ of this to be 0 we need $0=a_{11}\rho_{22}-\Re\left(a_{12}\rho_{21}\right)-a_{22}\rho_{22}$.
Now, for this to hold for every value of $a_{11},a_{22}$ and $a_{12}$ we
need $\rho_{22}=\rho_{21}=0$ leaving us with $\rho=\left(\begin{matrix}1 & 0\\
0 & 0
\end{matrix}\right)$
in which case 
\begin{equation}
\text{Tr}\left(A\rho\right)=a_{11}
\end{equation}

And so we cannot learn anything about $a_{12}$ or $a_{22}$ with
this initial state and so cannot learn all of $A$ without linear
order affect of the noise.

\subsection{Lemma - the result of the strong measurement with post-selection is
not affected by the noise in linear order only if the initial state
is not affected by the noise in linear order}

\begin{equation}
\left\langle Q\right\rangle _{strong\,with\,post}=\sum_{i}a_{i}\left\langle \psi_{f}|a_{i}\right\rangle \left\langle a_{i}|\rho|a_{i}\right\rangle \left\langle a_{i}|\psi_{f}\right\rangle 
\end{equation}
\begin{equation}
\left\langle Q\right\rangle _{noisy\,strong\,with\,post}=\sum_{i}a_{i}\left\langle \psi_{f}|a_{i}\right\rangle \left\langle a_{i}|\mathcal{E}\left(\rho\right)|a_{i}\right\rangle \left\langle a_{i}|\psi_{f}\right\rangle 
\end{equation}

We will now use the notation $\text{linear}\left(M\right)$
to denote the part in $M$ (for any $M$), which is
linear in the noise parameter $\gamma$:
\begin{align}
\Delta_{\mathcal{E}} & =\text{linear}\left(\left\langle Q\right\rangle _{noisy\,strong\,with\,post}-\left\langle Q\right\rangle _{strong\,with\,post}\right)\\
 & =\sum_{i}a_{i}\left\langle \psi_{f}|a_{i}\right\rangle \left\langle a_{i}|\text{linear}\left(\mathcal{E}\left(\rho\right)-\rho\right)|a_{i}\right\rangle \left\langle a_{i}|\psi_{f}\right\rangle \\
 & =\sum_{i}a_{i}\left\langle a_{i}|\text{linear}\left(\mathcal{E}\left(\rho\right)-\rho\right)|a_{i}\right\rangle \left|\left\langle a_{i}|\psi_{f}\right\rangle \right|^{2} \label{Delta}
\end{align}

We will now prove that the only case for which the linear term of the measurement error vanishes for all operators $A$ is when the initial state is not affected by the noise channel in the linear order. We will define $\mathcal{E}_{damping}$ as the family of amplitude and phase damping channels. More precisely, we need to prove the following:

\paragraph*{Lemma: }

$\forall \rho$ for which $\exists$ $\mathcal{E}\in\mathcal{E}_{damping}$ for which $\text{linear}\left(\mathcal{E}\left(\rho\right)-\rho\right)\neq 0$, $\not\exists |\psi_f\rangle$ for which $\Delta_{\mathcal{E}} = 0$ $\forall A$.  

We will do this using a proof-by-contradiction method, and thus we start by stating the contrapositive form of the above, which results in the following claim:

\subsubsection{Claim: }

For any fixed $|\psi_{f}\rangle$ and any fixed $\rho$ for which there exists $\mathcal{E}\in\mathcal{E}_{damping}$ for which
$\text{linear}\left(\mathcal{E}\left(\rho\right)-\rho\right)\neq0$,
there exists an operator $A$ for which $\Delta_{\mathcal{E}}\neq0$.

Indeed this Claim is equivalent to the lemma, and can be proven in a straightforward manner by showing a specific operator $A$ satisfying $\Delta_{\mathcal{E}}\neq0$.

\subsubsection{Proof of claim:}

Given some fixed $|\psi_{f}\rangle$ and fixed $\rho$ for which $\text{linear}\left(\mathcal{E}\left(\rho\right)-\rho\right)\neq0$,
we choose $A$ to be $A=|a\rangle\langle a|$ where $|a\rangle$ is
normalized, and will be defined shortly.
For such an operator $A$ we get that $\Delta_{\mathcal{E}}$ in Eq. \ref{Delta} obtains the form 
\begin{equation}
\Delta_{\mathcal{E}}=\left\langle a|\text{linear}\left(\mathcal{E}\left(\rho\right)-\rho\right)|a\right\rangle \left|\left\langle a|\psi_{f}\right\rangle \right|^{2}\label{eq:linear noise for general a}
\end{equation}

We will proceed by finding a vector $|a\rangle$ for which $\Delta_{\mathcal{E}}$ in Eq. (\ref{eq:linear noise for general a}) is not zero.

We will use the notation  $\sigma=\text{linear}\left(\mathcal{E}\left(\rho\right)-\rho\right)$.
We now prove that $\sigma$ is hermitian. We notice
that $\rho$ and $\mathcal{E}\left(\rho\right)$ are both quantum
states and so are hermitian, and hence $\mathcal{E}\left(\rho\right)-\rho$
is hermitian as well, since a sum of hermitian matrices is always hermitian.
Being hermitian, $\mathcal{E}\left(\rho\right)-\rho$ is unitarily
diagonalizable with real eigenvalues. We consider $\sigma=\text{linear}\left(\mathcal{E}\left(\rho\right)-\rho\right)$
in the basis that diagonalizes $\mathcal{E}\left(\rho\right)-\rho$.
Every non-diagonal element is zero in $\mathcal{E}\left(\rho\right)-\rho$
and so it is $0$ in $\sigma$ as well. And so $\sigma$ is unitarily
diagonalizable.
Moreover, the elements on the diagonal must be real. This is true since if there were non real elements on the diagonal of $\sigma$ higher order terms in the noise parameter cannot cancel linear order terms in the noise parameter, thus leading to non real terms on the diagonal of $\mathcal{E}\left(\rho\right)-\rho$, in contradiction to it being hermitian.
And so $\sigma$ is unitarily diagonalizable with real eigenvalues and so $\sigma$ is hermitian.

And so in the orthonormal basis that diagonalizes $\mathcal{E}\left(\rho\right)-\rho$ we can write $\sigma$ as
\begin{equation}
\sigma=s_{J}|v_{J}\rangle\langle v_{J}|+s_{I}|v_{I}\rangle\langle v_{I}|
\end{equation}
for orthonormal $\left\{ |v_{J}\rangle,|v_{I}\rangle\right\} $ and real $s_J, s_I$.
Since $\sigma=\text{linear}\left(\mathcal{E}\left(\rho\right)-\rho\right)\neq 0$ then at least one of $\left\{ s_{J},s_{I}\right\} $
is non-zero. Without loss of generality let us assume $s_{J}\neq0$.

We now divide into cases and choose $|a\rangle$ that gives $\Delta \neq 0$ for each case separately.

\paragraph{Case 1: }
$\left\langle \psi_{f}|\sigma|\psi_{f}\right\rangle \neq0$.

In this case, we can choose $|a\rangle=|\psi_{f}\rangle$ for which
Eq. (\ref{eq:linear noise for general a}) takes the form 
\begin{equation}
\Delta_{\mathcal{E}}=\left\langle a|\sigma|a\right\rangle \left|\left\langle a|\psi_{f}\right\rangle \right|^{2}=\left\langle \psi_{f}|\sigma|\psi_{f}\right\rangle \left|\left\langle \psi_{f}|\psi_{f}\right\rangle \right|^{2}=\left\langle \psi_{f}|\sigma|\psi_{f}\right\rangle \neq0.
\end{equation}

\paragraph{Case 2: }
$\left\langle \psi_{f}|\sigma|\psi_{f}\right\rangle =0$ and $\left\langle v_{J}|\psi_{f}\right\rangle =0$.

Since $|v_{J}\rangle$ and $|\psi_{f}\rangle$ are orthogonal in this
case, and they are both normalized, they form an orthonormal basis.
In this case we choose $|a\rangle=\frac{1}{\sqrt{2}}\left(|\psi_{f}\rangle+|v_{J}\rangle\right)$
for which 
\begin{equation}
\Delta_{\mathcal{E}}=\left\langle a|\sigma|a\right\rangle \left|\left\langle a|\psi_{f}\right\rangle \right|^{2}
\end{equation}
and
\begin{equation}
\left\langle a|\sigma|a\right\rangle =\frac{1}{2}\left(\left\langle \psi_{f}|\sigma|\psi_{f}\right\rangle +\left\langle v_{J}|\sigma|\psi_{f}\right\rangle +\left\langle \psi_{f}|\sigma|v_{J}\right\rangle +\left\langle v_{J}|\sigma|v_{J}\right\rangle \right).
\end{equation}
We will compute each of the terms separately.
In this case, by definition
\begin{equation}
\left\langle \psi_{f}|\sigma|\psi_{f}\right\rangle=0.
\end{equation}
The other terms are given by
\begin{equation}
\left\langle \psi_{f}|\sigma|v_{J}\right\rangle =\left\langle \psi_{f}|\left(s_{J}|v_{J}\rangle\langle v_{J}|+s_{I}|v_{I}\rangle\langle v_{I}|\right)|v_{J}\right\rangle =s_{J}\left\langle \psi_{f}|v_{J}\right\rangle \left\langle v_{J}|v_{J}\right\rangle +s_{I}\left\langle \psi_{f}|v_{I}\right\rangle \left\langle v_{I}|v_{J}\right\rangle =0
\end{equation}
and so since $\sigma$ is hermitian, also
\begin{equation}
\left\langle v_{J}|\sigma|\psi_{f}\right\rangle =0
\end{equation}
Lastly, by definition
\begin{equation}
\left\langle v_{J}|\sigma|v_{J}\right\rangle =s_{J}
\end{equation}
and so 
\begin{equation}
\left\langle a|\sigma|a\right\rangle =\frac{1}{2}\left(\left\langle \psi_{f}|\sigma|\psi_{f}\right\rangle +\left\langle v_{J}|\sigma|\psi_{f}\right\rangle +\left\langle \psi_{f}|\sigma|v_{J}\right\rangle +\left\langle v_{J}|\sigma|v_{J}\right\rangle \right)=\frac{1}{2}s_{J}\neq0
\end{equation}
\begin{equation}
\left|\left\langle a|\psi_{f}\right\rangle \right|^{2}=\left|\frac{1}{\sqrt{2}}\left\langle \psi_{f}|\psi_{f}\right\rangle \right|^{2}=\frac{1}{4}\neq0
\end{equation}
Therefore, Eq. (\ref{eq:linear noise for general a}) takes the form 
\begin{equation}
\Delta_{\mathcal{E}}=\left\langle a|\sigma|a\right\rangle \left|\left\langle a|\psi_{f}\right\rangle \right|^{2}=\frac{1}{8}s_{J}\neq0
\end{equation}

\paragraph{Case 3: }
$\left\langle \psi_{f}|\sigma|\psi_{f}\right\rangle =0$ and $\left\langle v_{J}|\psi_{f}\right\rangle \neq0$.

In this case we choose $|a\rangle=C\left(|v_{J}\rangle+\frac{1}{\left\langle v_{J}|\psi_{f}\right\rangle }|\psi_{f}\rangle\right)$
where $C$ is a normalization factor. In this case we get 
\begin{equation}
\left\langle \psi_{f}|\sigma|v_{J}\right\rangle =\left\langle \psi_{f}|\left(s_{J}|v_{J}\rangle\langle v_{J}|+s_{I}|v_{I}\rangle\langle v_{I}|\right)|v_{J}\right\rangle =s_{J}\left\langle \psi_{f}|v_{J}\right\rangle \left\langle v_{J}|v_{J}\right\rangle +s_{I}\left\langle \psi_{f}|v_{I}\right\rangle \left\langle v_{I}|v_{J}\right\rangle =s_{J}\left\langle \psi_{f}|v_{J}\right\rangle 
\end{equation}
and so 
\begin{equation}
\left\langle v_{J}|\sigma|\psi_{f}\right\rangle = s_{J}\left\langle v_{J}|\psi_{f}\right\rangle.
\end{equation}
While as in the previous case
\begin{equation}
\left\langle v_{J}|\sigma|v_{J}\right\rangle =s_{J}
\end{equation}
and
\begin{equation}
\left\langle \psi_{f}|\sigma|\psi_{f}\right\rangle=0.
\end{equation}
And so in total
\begin{align}
\left\langle a|\sigma|a\right\rangle  & =\left|C\right|^{2}\left(\left\langle v_{J}|\sigma|v_{J}\right\rangle +\frac{1}{\left\langle v_{J}|\psi_{f}\right\rangle }\left\langle v_{J}|\sigma|\psi_{f}\right\rangle +\frac{1}{\left\langle \psi_{f}|v_{J}\right\rangle }\left\langle \psi_{f}|\sigma|v_{J}\right\rangle +\frac{1}{\left|\left\langle v_{J}|\psi_{f}\right\rangle \right|^{2}}\left\langle \psi_{f}|\sigma|\psi_{f}\right\rangle \right)\\
 & =\left|C\right|^{2}\left(s_{J}+\frac{1}{\left\langle v_{J}|\psi_{f}\right\rangle }s_{J}\left\langle v_{J}|\psi_{f}\right\rangle +\frac{1}{\left\langle \psi_{f}|v_{J}\right\rangle }s_{J}\left\langle \psi_{f}|v_{J}\right\rangle \right)\\
 & =\left|C\right|^{2}\left(s_{J}+s_{J}+s_{J}\right)\\
 & =3\left|C\right|^{2}s_J\\
 & \neq0
\end{align}
\begin{equation}
\left|\left\langle a|\psi_{f}\right\rangle \right|^{2}=\left|C\right|^{2}\left|\left\langle v_{J}|\psi_{f}\right\rangle +\frac{\left\langle \psi_{f}|\psi_{f}\right\rangle }{\left\langle \psi_{f}|v_{J}\right\rangle }\right|^{2}=\left|\frac{C}{\left\langle \psi_{f}|v_{J}\right\rangle }\left(\left|\left\langle \psi_{f}|v_{J}\right\rangle \right|^{2}+1\right)\right|^{2}\neq0
\end{equation}
where the last inequality holds since $\left|\left\langle \psi_{f}|v_{J}\right\rangle \right|^{2}$
is not negative and so $\left(\left|\left\langle \psi_{f}|v_{J}\right\rangle \right|^{2}+1\right)$
is strictly positive. 
Therefore, Eq. (\ref{eq:linear noise for general a}) takes the form 
\begin{equation}
\Delta_{\mathcal{E}}=\left\langle a|\sigma|a\right\rangle \left|\left\langle a|\psi_{f}\right\rangle \right|^{2}=3\left|C\right|^{2}s_J\left|\frac{C}{\left\langle \psi_{f}|v_{J}\right\rangle }\left(\left|\left\langle \psi_{f}|v_{J}\right\rangle \right|^{2}+1\right)\right|^{2}\neq0
\end{equation}

Therefore, the linear order $\Delta$ vanishes only when the linear order
of the noise vanishes, i.e. $\text{linear}\left(\mathcal{E}\left(\rho\right)-\rho\right)=0$.

\subsection{Proof of Claim 3 - strong measurement with post-selection cannot accomplish
the task even for amplitude damping alone}
Following the lemma, we need to identify the initial states which
are not affected by amplitude and phase damping noise in the linear
order.

\begin{equation}
AD:\,E_{0}=\left(\begin{matrix}1 & 0\\
0 & \sqrt{1-\gamma}
\end{matrix}\right),\,\,E_{1}=\left(\begin{matrix}0 & \sqrt{\gamma}\\
0 & 0
\end{matrix}\right)
\end{equation}
\begin{equation}
PD:\,E_{0}=\left(\begin{matrix}1 & 0\\
0 & \sqrt{1-\lambda}
\end{matrix}\right),\,\,E_{1}=\left(\begin{matrix}0 & 0\\
0 & \sqrt{\lambda}
\end{matrix}\right)
\end{equation}
\begin{align}
\mathcal{E}_{AD}\left(\rho\right) & =\left(\begin{matrix}1 & 0\\
0 & \sqrt{1-\gamma}
\end{matrix}\right)\left(\begin{matrix}\rho_{11} & \rho_{12}\\
\rho_{21} & \rho_{22}
\end{matrix}\right)\left(\begin{matrix}1 & 0\\
0 & \sqrt{1-\gamma}
\end{matrix}\right)+\left(\begin{matrix}0 & \sqrt{\gamma}\\
0 & 0
\end{matrix}\right)\left(\begin{matrix}\rho_{11} & \rho_{12}\\
\rho_{21} & \rho_{22}
\end{matrix}\right)\left(\begin{matrix}0 & 0\\
\sqrt{\gamma} & 0
\end{matrix}\right)\\
 & =\left(\begin{matrix}1 & 0\\
0 & \sqrt{1-\gamma}
\end{matrix}\right)\left(\begin{matrix}\rho_{11} & \sqrt{1-\gamma}\rho_{12}\\
\rho_{21} & \sqrt{1-\gamma}\rho_{22}
\end{matrix}\right)+\left(\begin{matrix}0 & \sqrt{\gamma}\\
0 & 0
\end{matrix}\right)\left(\begin{matrix}\sqrt{\gamma}\rho_{12} & 0\\
\sqrt{\gamma}\rho_{22} & 0
\end{matrix}\right)\\
 & =\left(\begin{matrix}\rho_{11} & \sqrt{1-\gamma}\rho_{12}\\
\sqrt{1-\gamma}\rho_{21} & \left(1-\gamma\right)\rho_{22}
\end{matrix}\right)+\left(\begin{matrix}\gamma p_{22} & 0\\
0 & 0
\end{matrix}\right)\\
 & =\left(\begin{matrix}\rho_{11}+\gamma p_{22} & \sqrt{1-\gamma}\rho_{12}\\
\sqrt{1-\gamma}\rho_{21} & \left(1-\gamma\right)\rho_{22}
\end{matrix}\right)
\end{align}
And the linear order of the noise vanishes only when $\rho_{12}=\rho_{21}=\rho_{22}=0$.
In that case, we have
\begin{equation}
|\psi_{f}\rangle=\left(\begin{matrix}f_{0}\\
f_{1}
\end{matrix}\right),\,\,\,\rho=\left(\begin{matrix}1 & 0\\
0 & 0
\end{matrix}\right),\,\,\left|f_{0}\right|^{2}\neq0
\end{equation}
\begin{equation}
A=\sum_{i}a_{i}|a_{i}\rangle\langle a_{i}|
\end{equation}
\begin{equation}
a_{i}=\left(\begin{matrix}a_{i,0}\\
a_{i,1}
\end{matrix}\right)
\end{equation}
\begin{equation}
\left\langle \psi_{f}|a_{i}\right\rangle =f_{0}^{*}a_{i,0}+f_{1}^{*}a_{i,1}
\end{equation}
\begin{equation}
\left\langle a_{i}|\rho|a_{i}\right\rangle =\left(\begin{matrix}a_{i,0}^{*} & a_{i,1}^{*}\end{matrix}\right)\left(\begin{matrix}1 & 0\\
0 & 0
\end{matrix}\right)\left(\begin{matrix}a_{i,0}\\
a_{i,1}
\end{matrix}\right)=\left(\begin{matrix}a_{i,0}^{*} & a_{i,1}^{*}\end{matrix}\right)\left(\begin{matrix}a_{i,0} & 0\\
0 & 0
\end{matrix}\right)=\left|a_{i,0}\right|^{2}
\end{equation}
\begin{align}
\left\langle Q\right\rangle _{strong\,with\,post} & =\sum_{i}a_{i}\left\langle \psi_{f}|a_{i}\right\rangle \left\langle a_{i}|\rho|a_{i}\right\rangle \left\langle a_{i}|\psi_{f}\right\rangle \\
 & =\sum_{i}a_{i}\left|f_{0}^{*}a_{i,0}+f_{1}^{*}a_{i,1}\right|^{2}\left|a_{i,0}\right|^{2}
\end{align}
Let us assume 
\begin{equation}
\bar{a}_{0}=\left(\begin{matrix}a_{00}\\
a_{01}
\end{matrix}\right)=\left(\begin{matrix}1\\
0
\end{matrix}\right),\,\,\bar{a}_{1}=\left(\begin{matrix}a_{10}\\
a_{11}
\end{matrix}\right)=\left(\begin{matrix}0\\
1
\end{matrix}\right)
\end{equation}
and so 
\begin{equation}
A=a_{0}|a_{0}\rangle\langle a_{0}|+a_{1}|a_{1}\rangle\langle a_{1}|=\left(\begin{matrix}a_{0} & 0\\
0 & a_{1}
\end{matrix}\right)
\end{equation}
\begin{equation}
|\psi_{f}\rangle=\left(\begin{matrix}f_{0}\\
f_{1}
\end{matrix}\right)
\end{equation}
For this case 
\begin{align}
\left\langle Q\right\rangle _{strong\,with\,post} & =\sum_{i}a_{i}\left\langle \psi_{f}|a_{i}\right\rangle \left\langle a_{i}|\rho|a_{i}\right\rangle \left\langle a_{i}|\psi_{f}\right\rangle \\
 & =\sum_{i}a_{i}\left|f_{i}\right|^{2}\rho_{ii}\\
 & =a_{0}\left|f_{0}\right|^{2}\rho_{00}+a_{1}\left|f_{1}\right|^{2}\rho_{11}
\end{align}
But since we are in the case of $\rho=\left(\begin{matrix}1 & 0\\
0 & 0
\end{matrix}\right)$ we are left with 
\begin{equation}
\left\langle Q\right\rangle _{strong\,with\,post}=a_{0}\left|f_{0}\right|^{2}
\end{equation}
And so we cannot learn $a_{1}$ even in this very simplified case.

\section{Additional Theorem 1 - weak value advantage at learning $A$ under a Pauli noise channel \label{app: additional theorem 1}}

\subsection{High level}

\emph{Theorem 1.---}
(Advantage for Pauli noise compared to strong measurements) When $A$ is an unknown Hermitian operator acting on a primary system
of a single qubit which suffers from a Pauli noise channel, then
the WVMP has a noise sensitivity advantage compared to strong measurements, but not compared to strong measurements with postselection.

\emph{Proof outline.---}
Notice that this theorem is comprised of three separate claims. The first claim is the ability of the WVMP to accomplish the task, the second claim is the inability of the strong measurement to accomplish the task, and the third claim is the ability of the strong measurement with postselection to accomplish the task.
The proof of each of these claims is comprised of two steps. 
WVMP ability to accomplish the task:
a) We start by identifying the sets of initial and final states for which the WV is not affected by the noise to linear order.
b) We show that $A$ can be fully learned via the WVs of the sets of initial and final states found in the previous step. 
These two steps together prove that the WVMP can accomplish the task of learning $A$ with no linear order effect of the noise.
The inability of the strong measurement, and the ability of the strong measurement with postselection to accomplish the task is done in a similar fashion. 

\emph{Proof of Claim 1 ---}
First we show that in the case of Pauli noise, the first order error in the WV is given by
\footnotesize
\begin{equation}
\Delta_{\mathcal{E}_{\mathcal{P}}}=\sum_{\underset{\sigma\neq I}{\sigma\in P^{n}}}\lambda_{\sigma}\left(\frac{\left\langle \psi_{f}|A\sigma\rho_s\sigma|\psi_{f}\right\rangle }{\left\langle \psi_{f}|\rho_s|\psi_{f}\right\rangle }-\frac{\left\langle \psi_{f}|\sigma\rho_s\sigma|\psi_{f}\right\rangle }{\left\langle \psi_{f}|\rho_s|\psi_{f}\right\rangle }\frac{\left\langle \psi_{f}|A\rho_s|\psi_{f}\right\rangle }{\left\langle \psi_{f}|\rho_s|\psi_{f}\right\rangle }\right)
\end{equation}
\normalsize
In order to find the cases where $\Delta_{\mathcal{E}_{\mathcal{P}}}=0$ for any choice of the values of $\lambda_\sigma$ in the Pauli channel, we demand that each term in the sum vanishes individually for any $A$. 
We then find all the pairs of initial and
final states which satisfy these constraints, i.e. for which the WV is not affected by the Pauli noise channel in first order for any $A$.
The solutions are
\begin{equation}
\rho_{s}^{(1)}\left(r\right)=\left(\begin{matrix}\frac{1}{2} & r\\
r & \frac{1}{2}
\end{matrix}\right),\,\,|\psi_{f}\rangle^{(1)}=\frac{1}{\sqrt{2}}\left(\begin{matrix}1\\
1
\end{matrix}\right),\,\,r\neq-\frac{1}{2},
\end{equation}
\begin{equation}
\rho_{s}^{(2)}\left(r\right)=\left(\begin{matrix}r & 0\\
0 & 1-r
\end{matrix}\right),\,\,|\psi_{f}\rangle^{(2)}=\left(\begin{matrix}1\\
0
\end{matrix}\right),\,\,r\neq0,
\end{equation}
\begin{equation}
\rho_{s}^{(3)}\left(r\right)=\left(\begin{matrix}r & 0\\
0 & 1-r
\end{matrix}\right),\,\,|\psi_{f}\rangle^{(3)}=\left(\begin{matrix}0\\
1
\end{matrix}\right),\,\,r\neq1,
\end{equation}
\begin{equation}
\rho_{s}^{(4)}=\left(\begin{matrix}\frac{1}{2} & 0\\
0 & \frac{1}{2}
\end{matrix}\right),|\psi_{f}\left(\theta,\varphi\right)\rangle^{(4)}=\left(\begin{matrix}\cos\theta\\
\sin\theta e^{i\varphi}
\end{matrix}\right).
\end{equation}

Next we show that we can express any $a_{ij}$ as combinations of WVs on these pairs of states, thus showing that we can learn $A$ exactly up to first order. 

\emph{Proof of Claim 2 ---}
We show that when using only strong measurements the expectation
value of the strong measurement will not be affected to first order by the noise, only when its value is $\text{Tr}\left(A\right)$. Learning all other functions of the elements in $A$ will be sensitive to noise to first order in $\gamma$.

\emph{Proof of Claim 3 ---}
Lastly, we show that with an initial maximally mixed state and suitable final states, we can learn $A$ using strong measurements and postselection with no effect of the noise to first order.

\subsection{Claim 1 -- WMP can accomplish the task}

We notice that we can simply replace $\rho$ by $\mathcal{E}\left(\rho\right)$
in Eq. (\ref{eq:mixed initial}) to obtain the noisy weak value in the case of mixed initial state

\begin{equation}
A_{w,\mathcal{E}}=\frac{\left\langle \psi_f|A\mathcal{E}\left(\rho_s\right)|\psi_f\right\rangle }{\left\langle \psi_f|\mathcal{E}\left(\rho_s\right)|\psi_f\right\rangle }.\label{eq:weakval mixed initial}
\end{equation}

Notice that the Pauli channel is a specific instance of the general type of channel $\mathcal{E}_{\left\{ p_{i}\right\} }(\rho_s)=(1-p)\rho_s+\sum_{i}p_{i}E_{i}\rho_{s}E_{i}^{\dagger}$ for $p=\sum_{i}p_{i}$. For any such channel we can simplify the noisy weak value

\begin{align}
A_{w,\mathcal{E}_{\left\{ p_{i}\right\} }} & =\frac{\left\langle \psi_{f}|A\mathcal{E}\left(\rho_{s}\right)|\psi_{f}\right\rangle }{\left\langle \psi_{f}|\mathcal{E}\left(\rho_{s}\right)|\psi_{f}\right\rangle }\\
 & =\frac{\left\langle \psi_{f}|A\left(1-p\right)\rho_{s}+\sum_{i}p_{i}E_{i}\rho_{s}E_{i}^{\dagger}|\psi_{f}\right\rangle }{\left\langle \psi_{f}|\left(1-p\right)\rho_{s}+\sum_{i}p_{i}E_{i}\rho_{s}E_{i}^{\dagger}|\psi_{f}\right\rangle }\\
 & =\frac{\left\langle \psi_{f}|A\rho_{s}|\psi_{f}\right\rangle +\frac{\sum_{i}p_{i}}{1-p}\left\langle \psi_{f}|AE_{i}\rho_{s}E_{i}^{\dagger}|\psi_{f}\right\rangle }{\left\langle \psi_{f}|\rho_{s}|\psi_{f}\right\rangle \left(1+\frac{\sum_{i}p_{i}}{1-p}\frac{\left\langle \psi_{f}|E_{i}\rho_{s}E_{i}^{\dagger}|\psi_{f}\right\rangle }{\left\langle \psi_{f}|\rho_{s}|\psi_{f}\right\rangle }\right)}\\
 & =\frac{\left\langle \psi_{f}|A\rho_{s}|\psi_{f}\right\rangle }{\left\langle \psi_{f}|\rho_{s}|\psi_{f}\right\rangle }+\sum_{i}p_{i}\left(\frac{\left\langle \psi_{f}|AE_{i}\rho_{s}E_{i}^{\dagger}|\psi_{f}\right\rangle }{\left\langle \psi_{f}|\rho_{s}|\psi_{f}\right\rangle }-\frac{\left\langle \psi_{f}|A\rho_{s}|\psi_{f}\right\rangle }{\left\langle \psi_{f}|\rho_{s}|\psi_{f}\right\rangle }\frac{\left\langle \psi_{f}|E_{i}\rho_{s}E_{i}^{\dagger}|\psi_{f}\right\rangle }{\left\langle \psi_{f}|\rho_{s}|\psi_{f}\right\rangle }\right)+O\left(p_{i}p_{j},p_{i}^{2}\right),
\end{align}

and so in total

\begin{equation}
A_{w,\mathcal{E}_{\left\{ p_{i}\right\} }}=\frac{\left\langle \psi_{f}|A\mathcal{E}\left(\rho_{s}\right)|\psi_{f}\right\rangle }{\left\langle \psi_{f}|\mathcal{E}\left(\rho_{s}\right)|\psi_{f}\right\rangle }=A_{w}+\sum_{i=1}^{n}p_{i}\Delta_{E_i}+O\left(p_{i}p_{j},p_{i}^{2}\right)
\end{equation}
for
\begin{equation}
\Delta_{E_i}=\frac{\left\langle \psi_{f}|AE_{i}\rho_{s}E_{i}^{\dagger}|\psi_{f}\right\rangle }{\left\langle \psi_{f}|\rho_{s}|\psi_{f}\right\rangle }-\frac{\left\langle \psi_{f}|E_{i}\rho_{s}E_{i}^{\dagger}|\psi_{f}\right\rangle }{\left\langle \psi_{f}|\rho_{s}|\psi_{f}\right\rangle }\frac{\left\langle \psi_{f}|A\rho_{s}|\psi_{f}\right\rangle }{\left\langle \psi_{f}|\rho_{s}|\psi_{f}\right\rangle }.\label{eq:general first order definition}
\end{equation}

We now solve the equations $\Delta_{E_i}=0$ for all $E_i=\sigma\in \{X,Y,Z\}$ single qubit Paulis.
The pairs of initial and final staets for which the leading order $\Delta_{E_i}$ vanishes  for all Pauli errors are:

\begin{equation}
\rho_{s,1}=\left(\begin{matrix}\frac{1}{2} & r\\
r & \frac{1}{2}
\end{matrix}\right)\,r\neq-\frac{1}{2},|\psi_{f}\rangle_{1}=\left(\begin{matrix}\frac{1}{\sqrt{2}}\\
\frac{1}{\sqrt{2}}
\end{matrix}\right),\left\langle \psi_{f}|\rho_{s}|\psi_{f}\right\rangle =\frac{1}{2}+r\neq0,A_{w}=\frac{1}{2}\left(a_{11}+a_{12}+a_{21}+a_{22}\right),\label{eq:pauli_sol1}
\end{equation}

\begin{equation}
\rho_{s,2}=\left(\begin{matrix}\lambda & 0\\
0 & 1-\lambda
\end{matrix}\right)\,\lambda\neq0,|\psi_{f}\rangle_{2}=\left(\begin{matrix}1\\
0
\end{matrix}\right),\left\langle \psi_{f}|\rho_{s}|\psi_{f}\right\rangle =\lambda\neq0,A_{w}=a_{11},\label{eq:pauli_sol2}
\end{equation}

\begin{equation}
\rho_{s,3}=\left(\begin{matrix}\lambda & 0\\
0 & 1-\lambda
\end{matrix}\right)\,\lambda\neq1,|\psi_{f}\rangle_{3}=\left(\begin{matrix}0\\
1
\end{matrix}\right),\left\langle \psi_{f}|\rho_{s}|\psi_{f}\right\rangle =1-\lambda\neq0,A_{w}=a_{22},\label{eq:pauli_sol3}
\end{equation}
and
\begin{equation}
\rho_{s,4}=\left(\begin{matrix}\frac{1}{2} & 0\\
0 & \frac{1}{2}
\end{matrix}\right),|\psi_{f}\rangle_{4}=\left(\begin{matrix}f_{1}\\
f_{2}
\end{matrix}\right),\left|f_{1}\right|^{2}+\left|f_{2}\right|^{2}=1,\left\langle \psi_{f}|\rho_{s}|\psi_{f}\right\rangle =\frac{1}{2},A_{w}=a_{11}\left|f_{1}\right|^{2}+a_{12}f_{2}f_{1}^{*}+a_{21}f_{1}f_{2}^{*}+a_{22}\left|f_{2}\right|^{2}.\label{eq:pauli_sol4}
\end{equation}

We have established the sets of states for which the WVs are not affected by the noise in the linear order.
Now, in order to show that the WVs succeed in the task of fully learning $A$ without linear order effect of the noise, we need to show that $A$ can be fully learned by the WVs of these specific pairs of states we presented above.
We denote $A_{w}\left(\rho,|\psi_{f}\rangle\right)$ as the WV for initial state $\rho$ and final state $|\psi_{f}\rangle$, 
and indeed $a_{11}=A_{w}\left(\rho_{s,2},|\psi_{f}\rangle_{2}\right)$,
$a_{22}=A_{w}\left(\rho_{s,3},|\psi_{f}\rangle_{3}\right)$, $a_{12}+a_{21}=A_{w}\left(\rho_{s,1},|\psi_{f}\rangle_{1}\right)-A_{w}\left(\rho_{s,4},\frac{1}{\sqrt{2}}{1 \choose -1}\right)$
and $i\left(a_{12}-a_{21}\right)=A_{w}\left(\rho_{s,1},|\psi_{f}\rangle_{1}\right)-A_{w}\left(\rho_{s,4},\frac{1}{\sqrt{2}}{1 \choose -1}\right)-2\left(A_{w}\left(\rho_{s,1},|\psi_{f}\rangle_{1}\right)-A_{w}\left(\rho_{s,4},\frac{1}{\sqrt{2}}{1 \choose i}\right)\right)$.

We have presented here a few different sets of initial and final states that can be used to learn A with no linear order affect of Pauli noise. Note that as will be shown in Appendix \ref{app: additional theorem 2}, this can be done for any unital noise with an initial maximally mixed state. It is beneficial to present here the full sets of states that can be used to overcome any Pauli noise, and specifically to show that this family is broader than the initial mixed state alone, since these additional possibilities could have practical advantages, for example when attempting to perform some coherent sensing.

\subsection{Claim 2 -- Strong measurement cannot accomplish the task}

The expectation of a strong measurement is

\begin{equation}
\left\langle Q\right\rangle _{\rho_{p}}\simeq g\text{Tr}\left(A\sum_{k}E_{k}\rho_{s}E_{k}^{\dagger}\right).
\end{equation}

We want to understand how the expectation value behaves for general
initial state $\rho$, and see when the leading term in $p$ vanishes.
For the Pauli channel, where $\sigma\in\left\{ X,Y,Z\right\} $ we
have

\begin{equation}
\frac{1}{g}\left\langle Q\right\rangle _{\rho_{p}}\simeq\left(1-p\right)\text{Tr}\left(A\rho_{s}\right)+p\text{Tr}\left(A\sigma\rho_{s}\sigma\right).
\end{equation}

So for the leading term in $p$ to vanish we need $\text{Tr}\left(A\sigma\rho_{s}\sigma\right)=\text{Tr}\left(A\rho_{s}\right)$
for all $\sigma$ simultaneously. All these together imply $\rho=\frac{1}{2}I$.
So the only state for which the noise is suppressed to first order
is when $\rho=\frac{1}{2}I$, in which case we have 
\begin{equation}
\text{Tr}\left(A\rho\right)=\frac{1}{2}\text{Tr}\left(A\right)=\frac{1}{2}\left(a_{11}+a_{22}\right).
\end{equation}

\subsection{Claim 3 -- Strong measurement with post-selection can accomplish the task}

For strong measurement and post-selection we have
\begin{equation}
\left\langle Q\right\rangle _{strong\,with\,post}=\sum_{i}a_{i}\left\langle \psi_{f}|a_{i}\right\rangle \left\langle a_{i}|\rho|a_{i}\right\rangle \left\langle a_{i}|\psi_{f}\right\rangle 
\end{equation}
For an initial maximally mixed state $\rho\propto I$, we have $\mathcal{E}_{\text{pauli}}\left(\rho\right)=\rho$,
and so $\left\langle Q\right\rangle _{strong\,with\,post}$ will not
be affected by Pauli noise. In this case
\begin{align}
\left\langle Q\right\rangle _{strong\,with\,post} & =\sum_{i}a_{i}\left\langle \psi_{f}|a_{i}\right\rangle \left\langle a_{i}|I|a_{i}\right\rangle \left\langle a_{i}|\psi_{f}\right\rangle \\
 & =\sum_{i}a_{i}\left\langle \psi_{f}|a_{i}\right\rangle \left\langle a_{i}|\psi_{f}\right\rangle \\
 & =\left\langle \psi_{f}|A|\psi_{f}\right\rangle 
\end{align}
And so, the full $A=\left(\begin{matrix}a_{00} & a_{01}\\
a_{10} & a_{11}
\end{matrix}\right)$ can be learned in this way:
\begin{equation}
|\psi_{f}\rangle=\left(\begin{matrix}1\\
0
\end{matrix}\right)\longrightarrow\left\langle Q\right\rangle _{strong\,with\,post}=a_{00}
\end{equation}
\begin{equation}
|\psi_{f}\rangle=\left(\begin{matrix}0\\
1
\end{matrix}\right)\longrightarrow\left\langle Q\right\rangle _{strong\,with\,post}=a_{11}
\end{equation}
\begin{equation}
|\psi_{f}\rangle=\frac{1}{\sqrt{2}}\left(\begin{matrix}1\\
1
\end{matrix}\right)\longrightarrow\left\langle Q\right\rangle _{strong\,with\,post}\propto a_{00}+a_{11}+2\mathcal{R}\left(a_{01}\right)
\end{equation}
\begin{equation}
|\psi_{f}\rangle=\frac{1}{\sqrt{2}}\left(\begin{matrix}1\\
i
\end{matrix}\right)\longrightarrow\left\langle Q\right\rangle _{strong\,with\,post}\propto a_{00}+a_{11}-2i\mathcal{I}\left(a_{01}\right)
\end{equation}

\section{Additional Theorem 2 - weak value advantage at learning $A$ under a Unital noise channel \label{app: additional theorem 2}}

\subsection{High level}

We show that the WVMP is advantageous over strong measurements without postselection
when the noise channel is a unital channel. A unital channel is a
channel $\mathcal{E}_{\text{unital}}\left(\rho\right)$ for which the maximally mixed state $\frac{1}{d}I$ is a fixed point, i.e. $\mathcal{E}_{\text{unital}}\left(\frac{1}{d}I\right)=\frac{1}{d}I$.

Theorem 2: (Advantage for unital noise compared to strong measurements) When $A$ is an unknown Hermitian
operator acting on a primary system of a single qubit which suffers
from a unital noise channel, then
the WVMP has a noise sensitivity advantage
compared to strong measurements, but not compared to strong measurements with postselection.

The Pauli channel is a specific instance of a unital channel. 
Since the noise in Theorem 2 is more general, the possibility results of Theorem 1 follow from Theorem 2 and the impossibility result of Theorem 2 follows from Theorem 1, but neither of the theorems as a whole follow from the other.
Another benefit stemming from the proof of Theorem 1 is additional options for combinations of pre- and postselected quantum states.

\emph{Proof Outline.---} By definition, for every unital channel, when $\rho_s=\frac{1}{2}I$
then
\begin{equation}
A_{w,\mathcal{E}}=\frac{\left\langle \psi_{f}|A\mathcal{E}_{\text{unital}}\left(\rho_s\right)|\psi_{f}\right\rangle }{\left\langle \psi_{f}|\mathcal{E}_{\text{unital}}\left(\rho_s\right)|\psi_{f}\right\rangle }=\frac{\left\langle \psi_{f}|A\rho_s|\psi_{f}\right\rangle }{\left\langle \psi_{f}|\rho_s|\psi_{f}\right\rangle }=A_{w}.
\end{equation}

And so, if the initial state is the maximally mixed state, the WV
under a unital channel is the ideal WV. We show that, interestingly
enough, $A$ can be fully learned using WVMP with the initial state
being maximally mixed. 
This is in contrast with common approaches in quantum sensing and metrology which rely on coherence \cite{degen2017quantum}.

The impossibility of success with only strong measurements follows from the impossibility shown in Theorem 1 due to the fact that the Pauli channel is a specific instance of a unital channel.
The proof of success of strong measurements with postselection for unital channels follows the same structure of Theorem 1's proof.
$\blacksquare$

The advantage we proved in the previous two theorems compared to strong measurements is important since the setting of strong measurements without postselection is used, e.g. for extracting expectation values in tomography tasks  \cite{PhysRevA.107.042403, ma2016pure}. No less important is the lack of advantage compared to strong measurements with postselection, which classifies these cases as partial advantages only, where the weak measurements are not the source of the advantage.

Next we move to our third result, which shows that the WVMP is advantageous even compared to strong measurements with postselection
when the noise channel is amplitude and phase damping. The amplitude
damping channel was defined above 
and the phase damping channel is defined as $\mathcal{E}_{\text{PD}}\left(\rho\right)=E_{0}\rho E_{0}^{\dagger}+E_{1}\rho E_{1}^{\dagger}$
for $E_{0}=\left(\begin{matrix}1 & 0\\
0 & \sqrt{1-\gamma}
\end{matrix}\right)$ and $E_{1}=\left(\begin{matrix}0 & 0\\
0 & \sqrt{\gamma}
\end{matrix}\right)$. 
These two channels commute \cite{schwartzman2024modeling} and so the combined channel of amplitude and phase damping is given by applying one channel after the other, which we denote as
$\mathcal{E}_{\text{PD}}\circ\mathcal{E}_{\text{AD}}\left(\rho\right)$.

\subsection{Claim 1 -- WMP can accomplish the task}

The weak value under noise is given by 
\begin{equation}
A_{w,\mathcal{E}}=\frac{\left\langle \psi_{f}|A\mathcal{E}\left(\rho_s\right)|\psi_{f}\right\rangle }{\left\langle \psi_{f}|\mathcal{E}\left(\rho_s\right)|\psi_{f}\right\rangle }
\end{equation}

Now, if the noise channel is a unital channel, and the initial state
is the maximally mixed state $\rho=\frac{1}{d}I$ then $\mathcal{E}\left(\rho\right)=\rho$
and so $A_{w,\mathcal{E}}=A_{w}$ and the weak value is not affected
by the noise. Now for
$A=\left(\begin{matrix}a_{11} & a_{12}\\
a_{21} & a_{22}
\end{matrix}\right)$
and
$|\psi_{f}\rangle=\left(\begin{matrix}f_{1}\\
f_{2}
\end{matrix}\right)$,
the weak value is given by 
\begin{align}
A_{w} & =\frac{\left\langle \psi_{f}|A|\psi_{f}\right\rangle }{\left\langle \psi_{f}|\psi_{f}\right\rangle }\\
 & =\left\langle \psi_{f}|A|\psi_{f}\right\rangle \\
 & =\left|f_{1}\right|^{2}a_{11}+f_{1}f_{2}^{*}a_{21}+f_{1}^{*}f_{2}a_{12}+\left|f_{2}\right|^{2}a_{22}.
\end{align}
And for $|\psi_{f}\rangle=\left(\begin{matrix}f_{1}\\
f_{2}
\end{matrix}\right)=\left(\begin{matrix}\cos\theta\\
\sin\theta e^{i\varphi}
\end{matrix}\right)$ we have 
\begin{equation}
A_{w}=\cos^{2}\theta a_{11}+\frac{1}{2}\sin\left(2\theta\right)e^{-i\varphi}a_{21}+\frac{1}{2}\sin\left(2\theta\right)e^{i\varphi}a_{12}+\sin^{2}\theta a_{22}.
\end{equation}

And so we can choose the final state to be
\begin{equation}
|\psi_{f}\rangle=\left(\begin{matrix}1\\
0
\end{matrix}\right)\longrightarrow A_{w}=a_{11},
\end{equation}
\begin{equation}
|\psi_{f}\rangle=\left(\begin{matrix}0\\
1
\end{matrix}\right)\longrightarrow A_{w}=a_{22},
\end{equation}
\begin{equation}
|\psi_{f}\rangle=\frac{1}{\sqrt{2}}\left(\begin{matrix}1\\
1
\end{matrix}\right)\longrightarrow A_{w}=\frac{1}{2}\left(a_{11}+a_{22}+2\mathcal{R}\left(a_{12}\right)\right),
\end{equation}
and
\begin{equation}
|\psi_{f}\rangle=\frac{1}{\sqrt{2}}\left(\begin{matrix}1\\
i
\end{matrix}\right)\longrightarrow A_{w}=\frac{1}{2}\left(a_{11}+a_{22}-2\mathcal{I}\left(a_{12}\right)\right).
\end{equation}

Therefore, we can fully learn $A$.

\subsection{Claim 2 -- strong measurements cannot accomplish the task}
Follows immediately from Claim 2 of Theorem 1, since the Pauli noise channel is a specific instance of the unital noise channel.

\subsection{Claim 3 -- strong measurements with post-selection can accomplish the task}
Is identical to the proof of Claim 3 in Theorem 1, since the maximally mixed state is not affected by any unital channel.

\section{Proof of maximally mixed state uniqueness for the unitary noise channel \label{app:maximally mixed}}

The result in this case consists of a few different claims which we
will prove here separately. 

\subparagraph{Claim 1: }

For any choice $\left\{ \rho_{s},|\psi_{f}\rangle\right\} $ where
$\rho_{s}$ is not the maximally mixed state, there exists a unitary
for which $\Delta_{\mathcal{E}_{\text{unitary}}}\neq0$ and so $A_{w}-A_{w,\mathcal{E}_{\text{unitary}}}=O\left(\gamma\right)$.

\subparagraph{Proof of Claim 1:}

We will show that the solutions for Pauli noise are not a solution
for all unitaries. And any solution that is not a solution for the Paulis cannot be a solution for all Paulis, since all the Paulis are unitary.

\begin{equation}
\rho_{s,1}=\left(\begin{matrix}\frac{1}{2} & r\\
r & \frac{1}{2}
\end{matrix}\right)\,r\neq-\frac{1}{2},|\psi_{f}\rangle_{1}=\left(\begin{matrix}\frac{1}{\sqrt{2}}\\
\frac{1}{\sqrt{2}}
\end{matrix}\right),\left\langle \psi_{f}|\rho_{s}|\psi_{f}\right\rangle =\frac{1}{2}+r\neq0,
\end{equation}

\begin{equation}
\Delta_{\mathcal{E}_{\text{hadamard}},1}=\frac{1}{4}r\left(1+2r\right)\left(a_{11}-a_{12}+a_{21}+a_{22}\right),
\end{equation}

which does not vanish only when $\rho_{s}$ is the maximally mixed
state.

\begin{equation}
\rho_{s,2}=\left(\begin{matrix}r & 0\\
0 & 1-r
\end{matrix}\right)\,r\neq0,|\psi_{f}\rangle_{2}=\left(\begin{matrix}1\\
0
\end{matrix}\right),\left\langle \psi_{f}|\rho_{s}|\psi_{f}\right\rangle =r\neq0,
\end{equation}

\[
\Delta_{\mathcal{E}_{\text{hadamard}},2}=\frac{1}{2}a_{12}r\left(2r-1\right),
\]

which does not vanish only when $\rho_{s}$ is the maximally mixed
state.
\begin{equation}
\rho_{s,3}=\left(\begin{matrix}r & 0\\
0 & 1-r
\end{matrix}\right)\,r\neq1,|\psi_{f}\rangle_{3}=\left(\begin{matrix}0\\
1
\end{matrix}\right),\left\langle \psi_{f}|\rho_{s}|\psi_{f}\right\rangle =1-r\neq0,
\end{equation}

\[
\Delta_{\mathcal{E}_{\text{hadamard}},3}=\frac{1}{2}a_{21}\left(1-r\right)\left(2r-1\right),
\]

which does not vanish only when $\rho_{s}$ is the maximally mixed
state.

\subparagraph{Claim 2: }

Under the simplifying assumption that also $\rho_{s}=|\psi_{s}\rangle\langle\psi_{s}|$
is pure, 

$\mathbb{E}_{U\sim Haar}\left[\Delta_{\mathcal{E}_{\text{unitary}}}\right]=\frac{1}{2\left\langle \psi_{f}|\rho_{s}|\psi_{f}\right\rangle }\left(\left\langle A\right\rangle _{\psi_{f}}-\left\langle A_{w}\right\rangle \right)$.

\subparagraph{Proof of Claim 2:}

We start by defining 
\begin{equation}
\Delta_{\mathcal{E}_{\text{unitary}}}=\frac{\left\langle \psi_{f}|AU\rho_{s}U^{\dagger}|\psi_{f}\right\rangle }{\left\langle \psi_{f}|\rho_{s}|\psi_{f}\right\rangle }-\frac{\left\langle \psi_{f}|U\rho_{s}U^{\dagger}|\psi_{f}\right\rangle }{\left\langle \psi_{f}|\rho_{s}|\psi_{f}\right\rangle }\frac{\left\langle \psi_{f}|A\rho_{s}|\psi_{f}\right\rangle }{\left\langle \psi_{f}|\rho_{s}|\psi_{f}\right\rangle },
\end{equation}
and
\begin{equation}
x=\left\langle \psi_{f}|AU\rho_{s}U^{\dagger}|\psi_{f}\right\rangle \left\langle \psi_{f}|\rho_{s}|\psi_{f}\right\rangle -\left\langle \psi_{f}|U\rho_{s}U^{\dagger}|\psi_{f}\right\rangle \left\langle \psi_{f}|A\rho_{s}|\psi_{f}\right\rangle ,
\end{equation}

and so 
$\Delta_{\mathcal{E}_{\text{unitary}}}=\frac{x}{\left\langle \psi_{f}|\rho_{s}|\psi_{f}\right\rangle ^{2}}$,
and so $x=0$ together with $\left\langle \psi_{f}|\rho_{s}|\psi_{f}\right\rangle=0$ which means the initial and final states are not orthogonal if and only if $\Delta_{\mathcal{E}_{\text{unitary}}}=0$.
Now, for $\rho_{f}=|\psi_{f}\rangle\langle\psi_{f}|$ we have

\begin{equation}
\mathbb{E}\left(x\right)=\left\langle \psi_{f}|\rho_{s}|\psi_{f}\right\rangle \mathbb{E}\left[\text{Tr}\left(U\rho U^{\dagger}\rho_{f}A\right)\right]-\left\langle \psi_{f}|A\rho_{s}|\psi_{f}\right\rangle \mathbb{E}\left[\text{Tr}\left(U\rho U^{\dagger}\rho_{f}\right)\right].
\end{equation}

Using Weingarten identities \cite{gu2013moments} we have $\mathbb{E}\left[\text{Tr}\left(UB_{1}U^{\dagger}b_{1}\right)\right]=Wg^{U}\left(I,2\right)\text{Tr}\left(B_{1}\right)\text{Tr}\left(b_{1}\right)=\frac{1}{2}\text{Tr}\left(B_{1}\right)\text{Tr}\left(b_{1}\right)$
and so $\mathbb{E}\left[\text{Tr}\left(U\rho_{s}U^{\dagger}\rho_{f}A\right)\right]=\frac{1}{2}\left\langle \psi_{f}|A|\psi_{f}\right\rangle $
and $\mathbb{E}\left[\text{Tr}\left(U\rho_{s}U^{\dagger}\rho_{f}\right)\right]=\frac{1}{2}$
and so $\mathbb{E}\left(x\right)=\frac{1}{2}\left(\left\langle \psi_{f}|\rho_{s}|\psi_{f}\right\rangle \left\langle \psi_{f}|A|\psi_{f}\right\rangle -\left\langle \psi_{f}|A\rho_{s}|\psi_{f}\right\rangle \right)$
and
\begin{equation}
\mathbb{E}\left(\Delta_{\mathcal{E}_{\text{unitary}}}\right)=\frac{1}{2}\frac{1}{\left\langle \psi_{f}|\rho_{s}|\psi_{f}\right\rangle }\left(\left\langle A\right\rangle _{f}-A_{w}\right).
\end{equation}

So the only other case where this will vanish is when $A_{w}=\left\langle A\right\rangle _{\psi_{f}}$
i.e. where the weak value is exactly the expectation value. In a two
dimensional space this can only occur if $|s\rangle=|\psi_{f}\rangle$
or $\langle\psi_{f}|A=\lambda\langle\psi_{f}|$.

Scaling: If $\left\langle \psi_{f}|\rho_{s}|\psi_{f}\right\rangle $
is very small and $A_{w}$ is very large then $\mathbb{E}\left(\Delta_{\mathcal{E}_{\text{unitary}}}\right)\propto\frac{1}{\left\langle \psi_{f}|\rho_{s}|\psi_{f}\right\rangle }A_{w}$
grows much faster than $A_{w}$.

\subparagraph{Claim 3: }

Under the simplifying assumptions stated previously,
\begin{equation}
\text{Var}\left(\Delta_{\mathcal{E}_{\text{unitary}}}\right)=\frac{1}{12}\frac{1}{\left\langle \psi_{f}|\rho_{s}|\psi_{f}\right\rangle ^{2}}\left(2\text{Var}\left(A\right)_{\psi_{f}}+\left|\left\langle A\right\rangle _{\psi_{f}}-A_{w}\right|^{2}\right).
\end{equation}

\subparagraph{Proof of Claim 3:}

\begin{align}
&\mathbb{E}\left(\left|x\right|^{2}\right) \\& =\int dU\left|\left\langle \psi_{f}|AU\rho_{s}U^{\dagger}|\psi_{f}\right\rangle \left\langle \psi_{f}|\rho_{s}|\psi_{f}\right\rangle -\left\langle \psi_{f}|U\rho_{s}U^{\dagger}|\psi_{f}\right\rangle \left\langle \psi_{f}|A\rho_{s}|\psi_{f}\right\rangle \right|^{2}\\
 & =\left\langle \psi_{f}|\rho_{s}|\psi_{f}\right\rangle ^{2}\int dU\text{Tr}\left(U\rho_{s}U^{\dagger}\rho_{f}U\rho_{s}U^{\dagger}A\rho_{f}A\right)+\left|\left\langle \psi_{f}|A\rho_{s}|\psi_{f}\right\rangle \right|^{2}\int dU\text{Tr}\left(U\rho_{s}U^{\dagger}\rho_{f}U\rho_{s}U^{\dagger}\rho_{f}\right)\\
 & \,\,\,\,-\left\langle \psi_{f}|\rho_{s}|\psi_{f}\right\rangle \left\langle \psi_{f}|\rho_{s}A|\psi_{f}\right\rangle \int dU\text{Tr}\left(U\rho_{s}U^{\dagger}\rho_{f}U\rho_{s}U^{\dagger}\rho_{f}A\right)-\left\langle \psi_{f}|\rho_{s}|\psi_{f}\right\rangle \left\langle \psi_{f}|A\rho_{s}|\psi_{f}\right\rangle \int dU\text{Tr}\left(U\rho_{s}U^{\dagger}\rho_{f}U\rho_{s}U^{\dagger}A\rho_{f}\right).
\end{align}

And again using Weingarten identities \cite{gu2013moments} we have

\begin{equation}
\mathbb{E}\left[\text{Tr}\left(UB_{1}U^{\dagger}b_{1}UB_{2}U^{\dagger}b_{2}\cdots B_{n-1}U^{\dagger}b_{n-1}UB_{n}U^{\dagger}b_{n}\right)\right]=\sum_{\alpha,\beta\in S_{n}}Wg^{U}\left(\beta\alpha^{-1},N\right)\text{Tr}_{\beta^{-1}}\left(B_{1},\cdots,B_{n}\right)\text{Tr}_{\alpha\gamma_{n}}\left(b_{1},\cdots,b_{n}\right),
\end{equation}

where $\text{Tr}_{\pi}\left(X_{1},\cdots,X_{n}\right)=\prod_{C\in\mathcal{C}\left(\pi\right)}\text{Tr}\left(\prod_{j\in C}X_{j}\right)$
and $\gamma_{n}=\left(1,2,\cdots,n\right)\in S_{n}$ is the cyclic
permutation. And since the initial and final states are pure we have
\begin{align}
\int dU\text{Tr}\left(U\rho_{s}U^{\dagger}\rho_{f}U\rho_{s}U^{\dagger}A\rho_{f}A\right) & =\frac{1}{6}\left\langle \psi_{f}|A|\psi_{f}\right\rangle ^{2}+\frac{1}{6}\left\langle \psi_{f}|A^{2}|\psi_{f}\right\rangle \\
\int dU\text{Tr}\left(U\rho_{s}U^{\dagger}\rho_{f}U\rho_{s}U^{\dagger}\rho_{f}\right) & =\frac{1}{3}\\
\int dU\text{Tr}\left(U\rho_{s}U^{\dagger}\rho_{f}U\rho_{s}U^{\dagger}\rho_{f}A\right) & =\frac{1}{3}\left\langle \psi_{f}|A|\psi_{f}\right\rangle \\
\int dU\text{Tr}\left(U\rho_{s}U^{\dagger}\rho_{f}U\rho_{s}U^{\dagger}A\rho_{f}\right) & =\frac{1}{3}\left\langle \psi_{f}|A|\psi_{f}\right\rangle. 
\end{align}

And so in total we have $\mathbb{E}\left(\left|x\right|^{2}\right)=\frac{1}{6}\left\langle \psi_{f}|\rho_{s}|\psi_{f}\right\rangle ^{2}\left(\left\langle A\right\rangle _{\psi_{f}}^{2}+\left\langle A^{2}\right\rangle _{\psi_{f}}+2\left|A_{w}\right|^{2}-4\left\langle A\right\rangle _{\psi_{f}}\mathcal{R}\left(A_{w}\right)\right)$
and

\begin{align}
\mathbb{E}\left(\left|\Delta_{\mathcal{E}_{\text{unitary}}}\right|^{2}\right) & =\frac{1}{6}\frac{1}{\left\langle \psi_{f}|\rho_{s}|\psi_{f}\right\rangle ^{2}}\left(\left\langle A\right\rangle _{\psi_{f}}^{2}+\left\langle A^{2}\right\rangle _{\psi_{f}}+2\left|A_{w}\right|^{2}-4\left\langle A\right\rangle _{\psi_{f}}\mathcal{R}\left(A_{w}\right)\right).
\end{align}

And so 

\begin{align}
\text{Var}\left(\Delta_{\mathcal{E}_{\text{unitary}}}\right) & =\mathbb{E}\left(\left|\Delta_{\mathcal{E}_{\text{unitary}}}\right|^{2}\right)-\left|\mathbb{E}\left(\Delta_{\rho,\mathcal{F}}^{unitary}\right)\right|^{2}\\
 & =\frac{1}{12}\frac{1}{\left\langle \psi_{f}|\rho_{s}|\psi_{f}\right\rangle ^{2}}\left(2\left(\left\langle A^{2}\right\rangle _{\psi_{f}}-\left\langle A\right\rangle _{\psi_{f}}^{2}\right)+\left|\left\langle A\right\rangle _{\psi_{f}}-A_{w}\right|^{2}\right).
\end{align}

Now, when $A_{w}$ is amplified then $A_{w}$ is much larger than
$\left\langle A\right\rangle _{\psi_{f}}$ and so also $\text{Var}\left(A\right)_{\psi_{f}}<\left|\left\langle A\right\rangle _{\psi_{f}}-A_{w}\right|^{2}$
for which case $\text{Var}\left(\Delta_{\mathcal{E}_{\text{unitary}}}\right)<\left|\mathbb{E}\left(\Delta_{\mathcal{E}_{\text{unitary}}}\right)\right|^{2}$
and the probability of of $\Delta_{\rho,\mathcal{F}}^{unitary}$ vanishing
is very low.

\subparagraph{Claim 4: }

The probability of $\Delta_{\mathcal{E}_{\text{unitary}}}$ to be
zero for a Haar randomly sampled $U$ is bounded above by $\frac{1}{3}\frac{2\text{Var}\left(A\right)_{f}+\left|\left\langle A\right\rangle _{f}-A_{w}\right|^{2}}{\left(\left\langle A\right\rangle _{f}-A_{w}\right)^{2}}$.

\subparagraph{Proof of Claim 4:}

Denoting $X$ a random variable which takes the value of $\Delta_{\mathcal{E}_{\text{unitary}}}\left(U\right)$
for a $U$ sampled Haar randomly. Due to Chebyshev's inequality
$
Pr\left(\left|X-\mu\right|\geq k\Delta\right)\leq\frac{1}{k^{2}}
$
 for expectation value $\mu$ and variance $\Delta^{2}$. For $k\Delta=\mu$
we have $\frac{1}{k^{2}}=\frac{\Delta^{2}}{\mu^{2}}$, and so in our
case
\begin{align}
Pr\left(\left|X-\mu\right|\geq\mu\right) & \leq\frac{\Delta^{2}}{\mu^{2}}\\
 & =\frac{\frac{1}{12}\frac{1}{\left\langle f|\rho|f\right\rangle ^{2}}\left(2\text{Var}\left(A\right)_{f}+\left|\left\langle A\right\rangle _{f}-A_{w}\right|^{2}\right)}{\left(\frac{1}{2}\frac{1}{\left\langle f|\rho|f\right\rangle }\left(\left\langle A\right\rangle _{f}-A_{w}\right)\right)^{2}}\\
 & =\frac{1}{3}\frac{2\text{Var}\left(A\right)_{f}+\left|\left\langle A\right\rangle _{f}-A_{w}\right|^{2}}{\left(\left\langle A\right\rangle _{f}-A_{w}\right)^{2}}
\end{align}
and so
\begin{equation}
Pr\left(X=0\right)\leq Pr\left(\left|X-\mu\right|\geq\mu\right)=\frac{1}{3}\frac{2\text{Var}\left(A\right)_{f}+\left|\left\langle A\right\rangle _{f}-A_{w}\right|^{2}}{\left(\left\langle A\right\rangle _{f}-A_{w}\right)^{2}}.
\end{equation}

\end{document}